\documentclass[MS]{iitmdiss}
\usepackage{titlesec}
\usepackage{times}
\usepackage{t1enc}
\usepackage{float}
\usepackage{amsmath, amssymb}
\usepackage{mathrsfs}
\usepackage[autostyle]{csquotes}
\usepackage{longtable}
\usepackage{lipsum} 
\usepackage{amsfonts}
\usepackage{amstext}
\usepackage{threeparttable}
\usepackage{textcomp}
\usepackage{rotating}
\usepackage{lscape}
\usepackage{gensymb}
\usepackage{placeins}
\usepackage{graphicx}
\usepackage{subfigure}
\usepackage{epstopdf}
\usepackage{wrapfig}
\usepackage{tabularx}
\usepackage{multirow}
\usepackage{amsmath} 
\usepackage{makecell}
\usepackage{pifont}
\usepackage{hhline}
\usepackage{cellspace}
\usepackage{enumitem}
\usepackage{mwe}
\usepackage{hyperref}
\hypersetup{
	colorlinks=true, 
	linktoc=all,     
	linkcolor=blue,  
	urlcolor=cyan,
	citecolor=magenta,
}
\usepackage{adjustbox}
\usepackage{ltablex} 
\usepackage{inputenc}
\usepackage{color}
\usepackage{framed}
\usepackage{afterpage}
\usepackage{array}
\usepackage{multirow}
\usepackage{geometry}
\usepackage{capt-of}
\usepackage{soul}
\usepackage{booktabs}
\usepackage{afterpage}
\usepackage{bm}
\usepackage[usenames,dvipsnames,svgnames,table]{xcolor} 
\usepackage{tikz}
\usetikzlibrary{shapes.geometric, arrows}
\usetikzlibrary{positioning}
\usepackage{verbatim}

\begin{document}


\title{EXPERIMENTAL STUDY OF AEROSOL DEPOSITION IN DISTAL LUNG BRONCHIOLES}

\author{ARNAB KUMAR MALLIK}

\date{AUGUST 2020}
\department{APPLIED MECHANICS}

\maketitle

\certificate

\vspace*{0.5in}

\noindent This is to certify that the thesis titled {\bf EXPERIMENTAL STUDY OF AEROSOL DEPOSITION IN DISTAL LUNG BRONCHIOLES}, submitted by {\bf ARNAB KUMAR MALLIK (AM17S032)}, 
  to the Indian Institute of Technology, Madras, for
the award of the degree of {\bf Master of Science}, is a bona fide
record of the research work done by him under my supervision. The
contents of this thesis, in full or in parts, have not been submitted
to any other Institute or University for the award of any degree or
diploma.

\vspace*{1.5in}

\begin{singlespacing}
\hspace*{-0.25in}
\parbox{2.5in}{
\noindent {\bf Mahesh V. Panchagnula} \\
\noindent Research Guide \\ 
\noindent Professor \\
\noindent Dept. of Applied Mechanics\\
\noindent IIT Madras, 600 036 \\
} 
\end{singlespacing}

\vspace*{0.25in}
\noindent Place: Chennai\\
Date:  
\newpage
\begin{center}
 \vspace*{100mm}
\begin{huge} \textit{Dedicated to my parents, wife}\end{huge}\\ 
\begin{huge} \textit{\& family...}\end{huge}\\
\end{center}

\acknowledgements

I am highly indebted to \textit{Prof. A.K. Mishra} from Department of Chemistry and \textit{Prof. Anju Chadha} from Department of Biotechnology, IIT Madras for providing their lab facilities for some of the experiments. \\
I acknowledge \textit{Defense Research and Development Organization} (DRDO) for supporting me financially to earn my MS degree form IIT Madras.\\
I thank my General Test Committee members, \textit{Dr. Satyanarayanan Seshadri} and \textit{Dr. Manikandan Mathur} for their support.\\
I thank the \textit{Head}, Department of Applied Mechanics, IIT Madras for maintaining a healthy research environment in our department. I also acknowledge various helps of \textit{lab assistants} and \textit{other staffs} of our department.\\
I gratefully acknowledge my lab mates for their constant support and selfless co-operation throughout my MS degree program.
\par
Finally, I gratefully acknowledge my advisor, \textbf{\textit{Prof. Mahesh V. Panchagnula}}, who not only guided me for my thesis but also enlightened me with other dimensions of life that made me a better human. I pay my humble gratitude and deepest sense of respect to him.


\abstract

\noindent KEYWORDS: \hspace*{0.5em} \parbox[t]{4.4in}{Aerosol, distal lungs, Reynolds number, deposition model.}

\vspace*{22pt}

\noindent The deposition of micron particles finds importance in meteorology and several engineering applications such as deposition of dust in gas lines, carbon deposition in engine exhaust, designing effective air-cleaning systems and estimating deposition of inhaled drug or atmospheric pollutants to determine its consequences on human health. Although the existing literature on deposition in straight tubes is quite mature, an experimental study on deposition in micro capillaries with a wide ranges of $Re$ that models particle dynamics in lungs, is missing. The deposition of atmospheric pollutants and nebulized drugs in the lung depends on various biological factors such as flow properties, lung morphology, breathing patterns, particle properties, deposition mechanism, etc. To complicate matters, each breath manifests flows spanning a wide range of Reynolds numbers ($10^{-2} \leqslant Re \leqslant 10^3$) in various regions of the lung. 
\par
In this study, the deposition of nebulized aerosol was experimentally investigated in phantom bronchioles of diameters relevant to the $7^{th}$ to the $23^{rd}$ branching generations and over the entire range of $Re$ manifest during one breathing cycle. The aerosol fluid was loaded with boron doped carbon quantum dots as a fluorophore. An aerosol was generated of this mixture fluid using an ultrasonic nebulizer, producing droplets of $6.5 \mu m$ as the mean diameter. The amount of aerosol deposited on the bronchiole walls was measured using a spectrofluorometer. The experimental results show that dimensionless deposition ($\delta$) varies inversely with the bronchiole aspect ratio ($\overline{L}$), with the effect of Reynolds number ($Re$) being significant only at low $\overline{L}$. $\delta$ also increased with increasing dimensionless bronchiole diameter ($\overline{D}$). However, it is invariant with the particle size based Reynolds number. It was observed that deposition can be distinguished into three distinct regimes based on $Re$- impaction dominated deposition (for $10 \leqslant Re \leqslant 10^3$), sedimentation dominated deposition (for $1 < Re < 10$) and diffusion dominated deposition (for $10^{-2} \leqslant Re \leqslant 1$). We also show that $\delta \overline{L} \sim Re^{-2}$ for $10^{-2} \leqslant Re \leqslant 1$, which is typical of diffusion dominated regime. For $Re>1$, $\delta \overline{L}$ is independent of $Re$, which is typical of impaction dominated regime. We also show a crossover regime where sedimentation becomes important. Finally, a universal bronchiole scale deposition model is proposed which can form the building block for lung-scale aerosol deposition prediction.
\par
The effect of breathing frequency and breath hold time on regional lung deposition is estimated from the experimentally developed model. Lower breathing frequency and higher breath hold time magnifies the deposition in the alveolar region which is highly desirable for efficient drug delivery. But for atmospheric pollutants and virus laden droplets, the low breathing rate accompanied with high breath hold time can highly affect human health.
\pagebreak


\begin{singlespace}
\tableofcontents
\thispagestyle{empty}

\listoftables
\addcontentsline{toc}{chapter}{LIST OF TABLES}
\listoffigures
\addcontentsline{toc}{chapter}{LIST OF FIGURES}
\end{singlespace}

\abbreviations

\noindent 
\begin{tabbing}
xxxxxxxxxxx \= xxxxxxxxxxxxxxxxxxxxxxxxxxxxxxxxxxxxxxxxxxxxxxxx \kill
\textbf{COPD}   \> Chronic Obstructive Pulmonary Disease \\
\textbf{WHO} \> World Health Organisation \\
\textbf{DALY}   \>  Disability Adjusted Life Years \\
\textbf{pMDI} \> Pressurised Meter Dose Inhaler \\
\textbf{DPI}   \> Dry Powder Inhaler \\
\textbf{PIV} \> Particle Image Velocimetry \\
\textbf{PTFE}   \>  Polytetrafluoroethylene \\
\textbf{CD} \> Carbon Quantum Dots \\
\textbf{QY}   \> Quantum Yield \\
\textbf{BCD} \> Boron doped Carbon Quantum Dots \\
\textbf{TEM}   \>  Transmission Electron Microscopy \\
\textbf{PDPA} \> Phase Doppler Particle Analyzer \\
\textbf{PMT}   \> Photo Multiplier Tube \\
\textbf{JICF}   \> Jet in Cross Flow \\
\textbf{\textit{pdf}}   \> Probability Density Function \\

\end{tabbing}

\pagebreak


\chapter*{\centerline{NOTATION}}
\addcontentsline{toc}{chapter}{NOTATION}

\begin{singlespace}
\begin{tabbing}
xxxxxxxxxxx \= xxxxxxxxxxxxxxxxxxxxxxxxxxxxxxxxxxxxxxxxxxxxxxxx \kill
\textbf{$G$}  \> Lung Generation \\
\textbf{$d$}   \> Measured aerosol deposition, ($ml$) \\
\textbf{$L$}  \> Length of bronchiole, ($mm$)  \\
\textbf{$D$}  \> Bronchiole diameter, ($mm$) \\
\textbf{$Q$}   \> Volume flow rate, ($ml/s$) \\
\textbf{$T$}  \> Time duration of the flow, ($s$) \\
\textbf{$\nu$}  \> Kinematic viscosity of air, ($m^2/s$) \\
\textbf{$D_{10}$}   \> Mean droplet diameter, ($\mu m$) \\
\textbf{$\alpha_0$}   \>  Volume fraction of aerosol from nebulizer, ($ml$ of aerosol per $ml$ of space) \\
\textbf{$D_F$}  \> Deposition Fraction \\
\textbf{$\overline{D}_F$}  \> $D_F$ per capillary length per unit time \\
\textbf{$\overline{L}$}  \> Aspect ratio \\
\textbf{$\overline{D}$}   \> Dimensionless bronchiole diameter \\
\textbf{$\overline{T}$}  \> Dimensionless Time \\
\textbf{$Re$}  \> Flow Reynolds number \\
\textbf{$Re_p$}   \> Particle Reynolds number \\
\textbf{$\delta$}  \> Deposition fraction per dimensionless length per dimensionless time
\end{tabbing}
\end{singlespace}

\pagebreak
\clearpage

\pagenumbering{arabic}


\chapter{INTRODUCTION}
\label{chap:intro}
The transport and deposition of aerosols in human lungs has attracted enormous attention from modern researchers for its application to pulmonary drug delivery and to minimise the adverse effect of atmospheric pollutants on human health. This chapter discusses the morphology of human lungs and breathing mechanisms, along with the transportation mechanics of aerosol particles in the lung airways. The objective and the scope of the study is discussed by highlighting the motivation for this study. Lastly, the overview of this thesis is presented in this chapter.

\section{Lung Morphology}

\begin{figure}[hbt!]
\centering
\includegraphics[width=11cm,height=14cm]{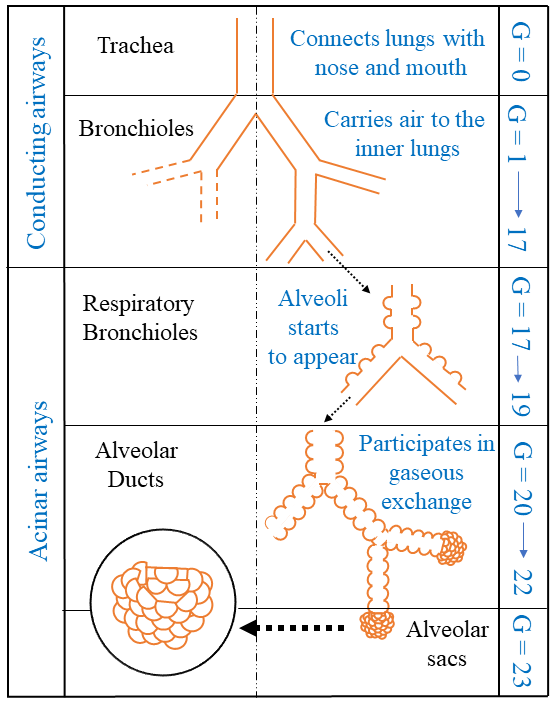}
\caption{Schematic of human lung morphology  proposed by \cite{weibel1963morphometry}. The schematic shows the conducting and acinar airways. The conducting airways transports air from nose and mouth to lungs through trachea whereas the acinar airways consists of alveolar ducts that participates in gaseous exchange with blood.}
\label{Figure: Lung Schematic}
\end{figure}

The human respiratory system comprises of (i) \textit{The upper respiratory tract} consisting of the nose, nasal passages, mouth, larynx, pharynx and trachea and (ii) \textit{The lower respiratory tract}, consisting of lungs. The human pulmonary system consists of left and right lungs, divided into slightly unequal proportions and connected to the upper respiratory system by the trachea. The architecture of the human lungs consists of a branching network through a sequences of bifurcation, known as bronchi. The diameter and length of the bronchi changes at each level of bifurcation, known as generation (G), starting with trachea as G = 0. A model for the morphological structure of the lung was proposed by \cite{weibel1963morphometry} where he presented a qualitative and quantitative description of the bronchial tree as shown in Figure \ref{Figure: Lung Schematic}, consisting of 23 levels of bifurcation or generations. 
Thus each half of the lungs has $2^{23} = 8388608$ airway tubes. The trachea (G = 0) supplies air to these $2^{23}$ ducts from the nose and mouth. The airways in the tracheobronchial region, i.e. $G = 0 \longrightarrow 16$, only conduct the flow of gases in and out from the lungs and do not participate in the gaseous exchange between the bloodstream and air. Thus, they constitute the anatomic dead space, approximately of volume 150 $ml$. In the pulmonary region, i.e. $G = 17 \longrightarrow 19$, the wall of airways are lined with air-exchange sacs, known as alveoli, which are the respiratory bronchioles. The alveoli are facilitated with capillary blood supply and can exchange gases between the inhaled air and bloodstream. For generations $G = 20 \longrightarrow 23$, known as the respiratory zone, the airways are completely made up of alveoli where G = 23 consists of clusters of alveoli. This zone consists of most of the lung volume which is about 2.5 to 3 $liters$. There are about $300 \times 10^6$ alveoli each of having diameter of $\sim$ 300 $\mu m$ in each human lung \citep{west2012respiratory}. Considering the alveoli to be of spherical shape the total surface area is estimated to be around $\sim $ 85 $m^2$ which is more than the area of a badminton court although their total volume is only about $4 l$. While $G = 0 \longrightarrow 16$ participate in conduction of air to lungs, $G = 17 \longrightarrow 23$ participate in exchanges of gases between blood and inhaled air \citep{grotberg2011respiratory}.

\section{Breathing Mechanism}
\begin{figure}[hbt!]
\centering
\includegraphics[width=12.5cm,height=11cm]{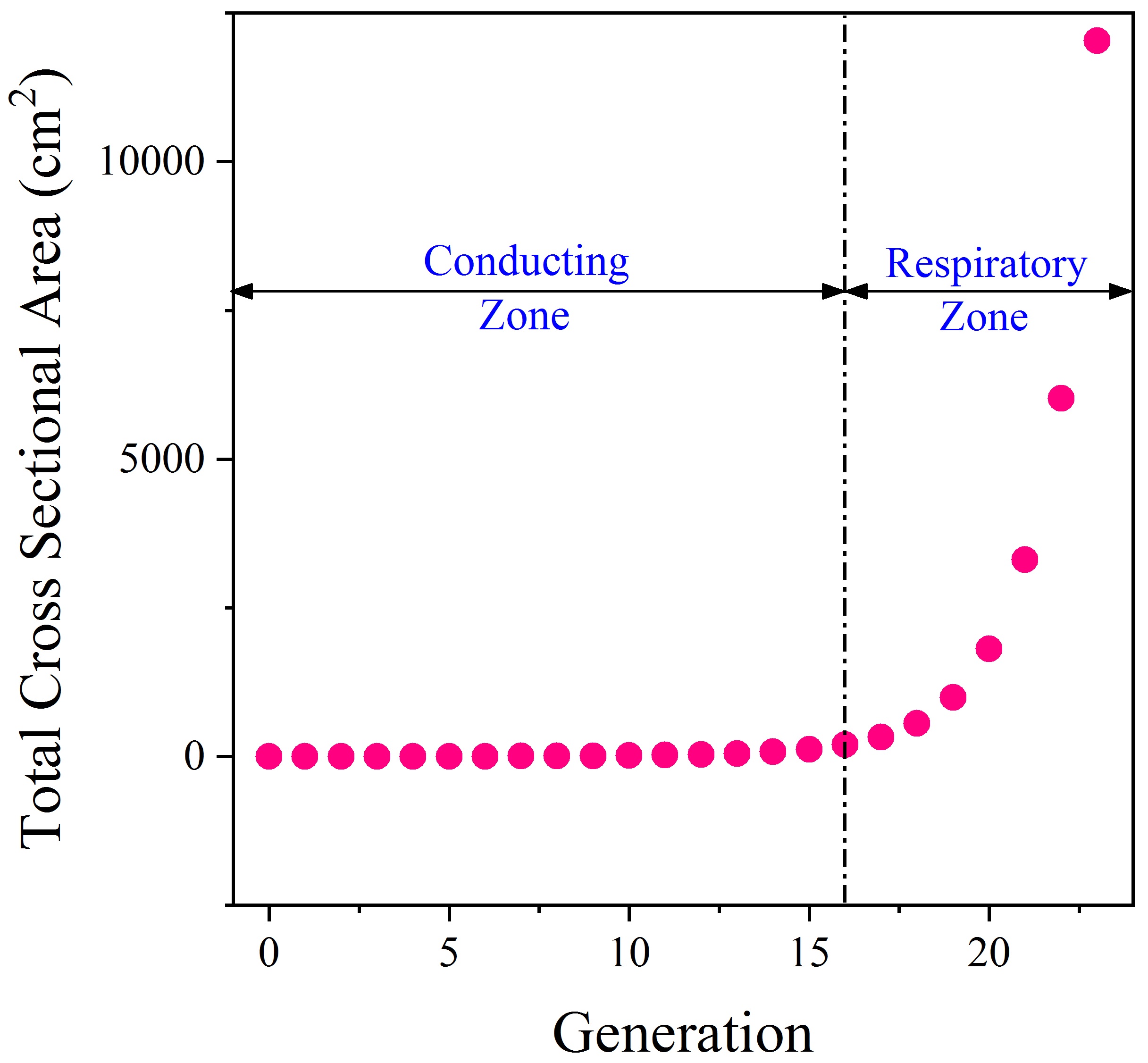}
\caption{Plot showing the variation of the total cross sectional area of the lung airways for different generations. The diameter of the airways decreases continuously as the number of generation increases, but the branching of airways at each generation causes the total cross sectional area to increase enormously for G > 16, which is identified as \textit{respiratory zone}.}
\label{Figure: C/s of airways}
\end{figure}

The respiration process is governed by the expansion and contraction of the diaphragm and the intercostal muscles. At the time of inspiration, the contraction of diaphragm raises the ribs, thus increasing the volume of the thoracic cavity as a result of which air is drawn into the lungs. The inspired air flows through the airways, up to the terminal bronchioles (G = 16) without any gaseous exchange with blood. Beyond G = 16, the total cross section area of the airways increases enormously (ref. Figure \ref{Figure: C/s of airways}) due to the branching of bronchioles and the velocity of air becomes very small.
In the last few generations, the air velocity becomes so less that diffusional mechanism becomes dominant. The tidal volume of human lungs is about 500 $ml$ which requires a low distending pressure of approximately 3 $millibar$, which is about 10 times lesser than the pressure required to inflate a rubber balloon of the same volume.

\section{Transport of Aerosol Particles in Lungs}
\begin{figure}[hbt!]
\centering
\includegraphics[width=12.5cm,height=11cm]{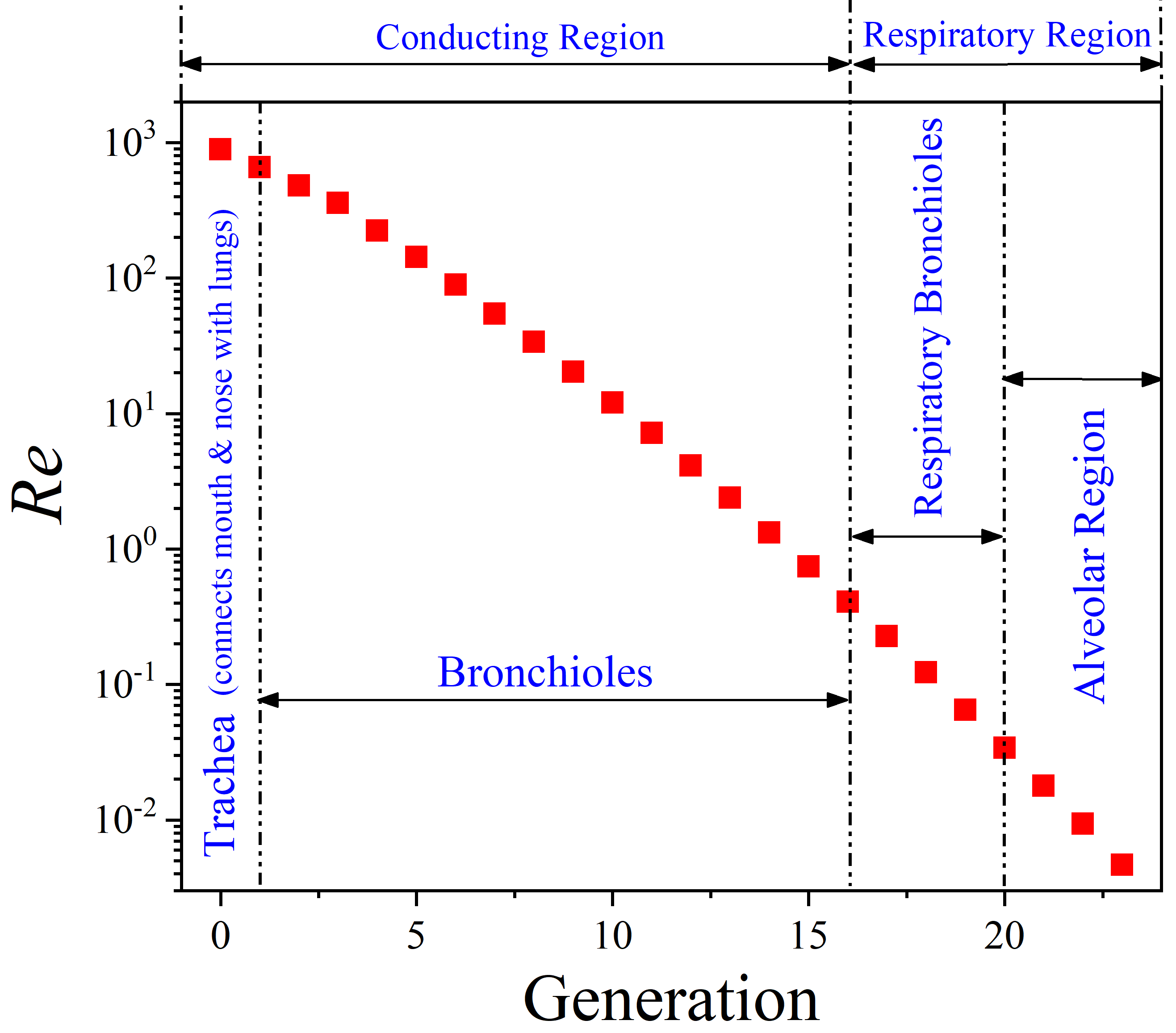}
\caption{Plot showing the variation of Reynolds number ($Re$) for different generations. Considering tidal volume of the lungs to be 500 $ml$ and one breathing cycle to be of 4 $sec$ (inspiration: 2 $sec$; expiration: 2 $sec$), the $Re$ is $\sim \mathcal{O}(10^3)$ in trachea and $\sim \mathcal{O}(10^{-1})$ in alveoli.}
\label{Figure: Re of different gen}
\end{figure}

The length and diameter of the airways varies enormously from generation to generation in the lung, starting with tracheal diameter of 18 $mm$ to 0.41 $mm$ at alveoli, as measured by \cite{weibel1963morphometry}. The lung dimensions are given in Table \ref{tab:lung dimension}. 
\begin{table}[htbp]
  \caption{Dimension of each generations of lungs \citep{weibel1963morphometry}}
  \begin{center}
  \begin{tabular}[c]{|c|c|c|c|} \hline
Genetations & Number of Branches & Diameter & Length \\ 
(G)         &        $2^G$       &  (cm)    &  (cm)  \\\hline
0 &	    1       &	  1.8    & 12  \\
1 &	    2       &	  1.22 	 & 4.76 \\
2 &	    4       &	  0.83 	 & 1.9  \\
3 &	    8       &	  0.56 	 & 1.76  \\
4 &	    16 	    &     0.45	 & 1.27  \\
5 &	    32      &	  0.35	 & 1.07  \\
6 &	    64      &	  0.28	 & 0.9 \\
7 &	    128     &	  0.23	 & 0.76 \\
8 &	    256	    &     0.186  &  0.64 \\
9 &	    512	    &     0.154  & 0.54 \\
10 &	1024    &	  0.13	 & 0.46 \\
11 &	2048    &	  0.109	 & 0.39 \\
12 &	4096    &	  0.095	 & 0.33 \\
13 &	8192    &	  0.082  & 0.27 \\
14 &	16384   &	  0.074	 & 0.23 \\
15 &	32768   &	  0.066	 & 0.2   \\
16 &	65536   &	  0.06	 & 0.165  \\
17 &	131072  &	  0.054	 & 0.141  \\
18 &	262144  &	  0.05	 & 0.117 \\
19 &	524288  &	  0.047	 & 0.099 \\
20 &	1048576 &	  0.045	 & 0.083 \\
21 &	2097152 &	  0.043	 & 0.07  \\
22 &	4194304 &	  0.041	 & 0.059  \\
23 &	8388608	&     0.041	 & 0.05 \\ \hline
  \end{tabular}
  \label{tab:lung dimension}
  \end{center}
\end{table}
This variation of airway diameter by about 4 orders of magnitude ($\sim O(10^2)$ at G = 0; $\sim O(10^{-1})$ at G = 23), causes a huge variation of air velocity an thus the Reynolds number ($Re$). The order of $Re$ varies several orders of magnitude over different generations as shown in Figure \ref{Figure: Re of different gen}, starting from $\mathcal{O}(10^3)$ in trachea to $\mathcal{O}(10^{-2})$ in alveoli, based on tidal volume of lungs to be 500 ml per inspiration of 2 seconds and kinematic viscosity of air at STP.
This variation of air velocity gives rise to several deposition mechanism, such as deposition by \textit{impaction}, \textit{sedimentation} and \textit{diffusion}. For very high $Re$ and large particles, the particles are offset from the air particle trajectory due to inertia and hit the wall of the airways which causes deposition due to \textit{impaction}. This kind of deposition mainly at the upper generation of lungs where $10 \leqslant Re \leqslant 10^3$. The deposition by \textit{sedimentation} occurs where $Re$ is relatively low, $1 < Re < 10$, when the gravitational force dominates over the inertia of the particles. \textit{Diffusion} occurs mainly due to Brownian motion of the particles at very low $Re$, ($10^{-2} \leqslant Re \leqslant 1$) and for small particles. The diffusion dominated deposition in the last few generation of lungs where the airway passages are extremely small and air flow velocity is negligible. 
\par
 Extrathoracic deposition is also a major cause of particle loss in the pulmonary system. The 90$^{\circ}$ bend geometry of the mouth and throat causes major fraction of the nebulized drug to deposit in that region, although the large particles are filtered out in the nose. The deposition of the smaller particles occurs in the conducting airways and are removed by a moving staircase of mucus that sweeps up the deposited particles to the epiglottis, where it is swallowed. The particles that deposit in the alveoli are engulfed by large wandering cells called macrophages. The flow of blood removes the foreign materials from the lungs and blood cells and poses a defense to them \cite{west2012respiratory}.

\section{Motivation}

Death that occurs by infection in the lower respiratory system, chronic obstructive pulmonary disease (COPD) and tracheal / bronchial cancer comprise of 17\% of the global death (Source: \href{https://www.who.int/healthinfo/global_burden_disease/about/en/}{World Health Organisation; WHO} ). The Global Burden of Disease study reported 251 million cases of COPD globally during the year 2016 which makes such respiratory diseases a matter of great concern. 
\begin{figure}[hbt!]
\centering
\includegraphics[width=12.3cm,height=11cm]{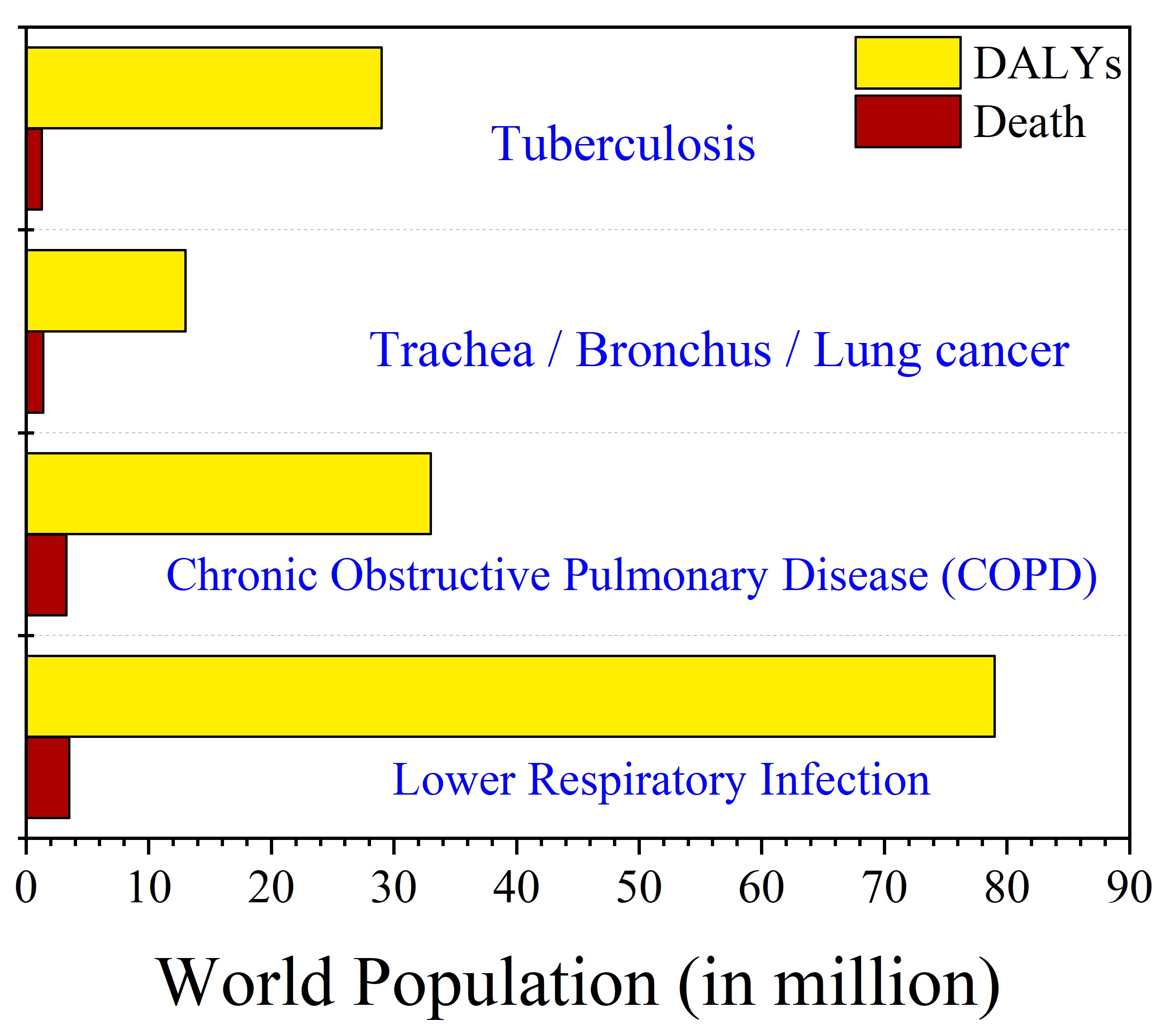}
\caption{Plot showing the death and disability adjusted life years of the global population due to respiratory diseases like tuberculosis, trachea / bronchus / lung cancer, chronic obstructive pulmonary disease (COPD), lower respiratory infections. The figure shows that about 9.5 million peoples died and 154 million people suffering from DALYs due to these respiratory diseases (report till 2008; source: \href{https://www.who.int/whosis/whostat/2011/en/}{World Heath Statistics 2011 by WHO}).}
\label{Figure: Global respiratory disease}
\end{figure}
The effect of respiratory diseases on global population is shown in Figure \ref{Figure: Global respiratory disease}, which caused death of 9.5 millions and disability adjusted life years (DALYs) of 154 million (report till 2008; source: \href{https://www.who.int/whosis/whostat/2011/en/}{World Heath Statistics 2011 by WHO}). For the period of 1990 - 2017, the average death rate due to chronic respiratory disease in India is about 90.5 per million, much higher than the global death rate which is around 55.03 per million (Ref: Figure \ref{Figure: COPD India vs world}; source: \href{https://gbd2016.healthdata.org/gbd-compare/india}{Global Burden of Diseases}). Investigation of aerosol deposition and particle transport in lungs can pave the way for health risk assessment due to the presence of atmospheric particulate matter and to asses the effectiveness of systemic drug delivery which is the main motivation behind this study. 
\par
The lung airways are often used as a route for treatment of various respiratory diseases since it provides rapid and non-invasive drug transfer to blood stream. The conventional treatment of lung disease rely on nebulized drug in the form of powder or liquid drop, which is targeted to the pulmonary and alveolar region of lungs since their large surface area provides rapid mixing of drug with the blood and avoids first pass metabolism. Thus, estimation of particle deposition helps in gauging the effectiveness of the drug delivery along with assessing the spreading of infections through inhalation.
\begin{figure}[hbt!]
\centering
\includegraphics[width=15cm,height=11cm]{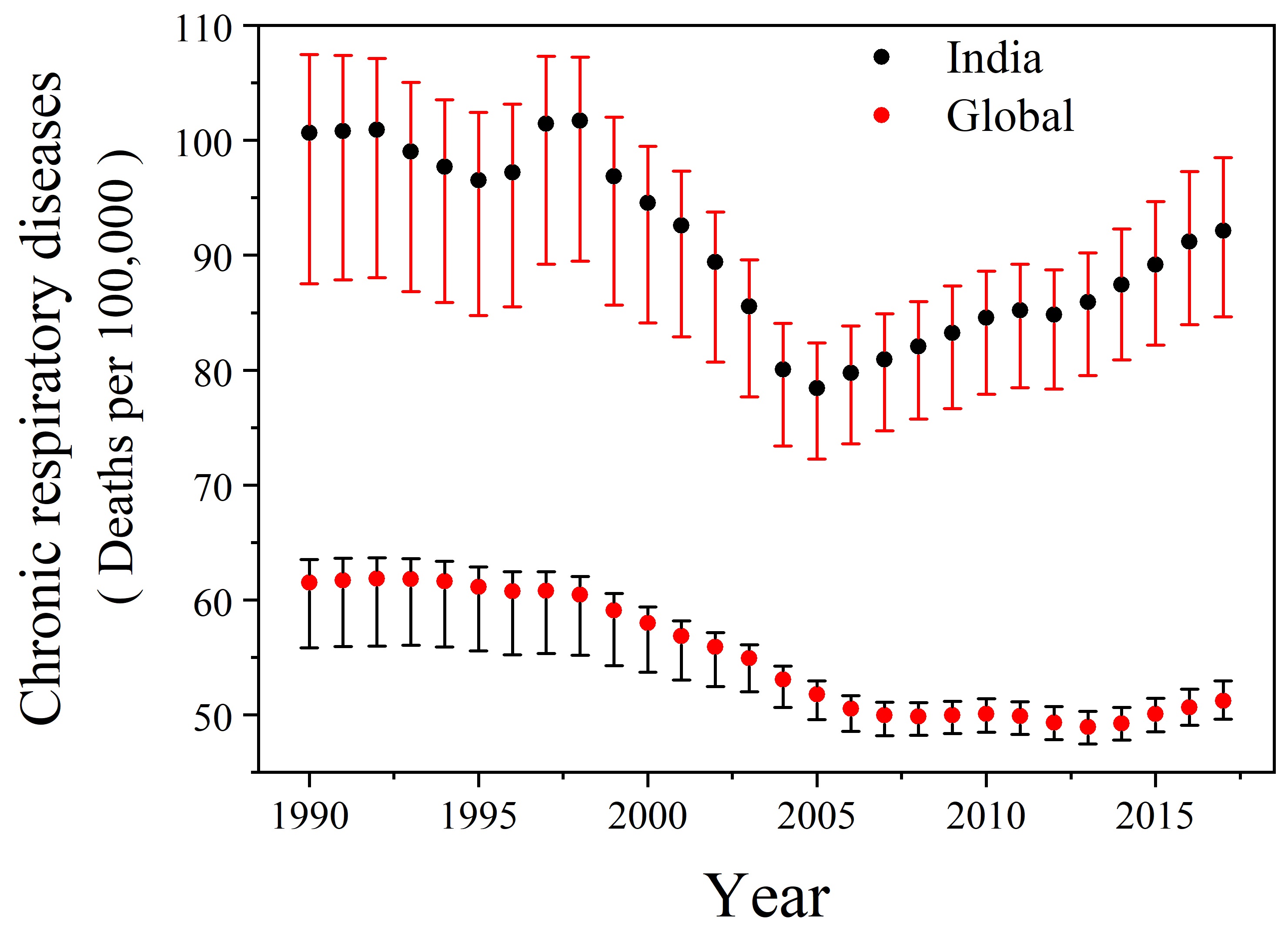}
\caption{Plot showing the death rate due to chronic respiratory diseases in India and world wide for the period of 1990 - 2017. The average death rate in India is about 90.5 per 100,000 which is much higher than the global death rate of 55.03 per million (Source: \href{https://gbd2016.healthdata.org/gbd-compare/india}{Global Burden of Diseases})}
\label{Figure: COPD India vs world}
\end{figure}
\par
Despite significant progress in quantifying particle deposition, both experimentally and numerically, there are still several discrepancies among the researchers regarding the fraction of the inhaled drug that actually reaches the pulmonary region. The drug delivery efficiencies for pressurised meter dose inhaler (pMDI) and dry powder inhalers (DPI) is approximately 40\% - 60\% \citep{clark2004pulmonary}, where the efficiencies for soft mist inhalers increase up to 80\%, although their pulmonary deposition is not reported. According to \cite{kleinstreuer2010airflow} the delivery efficiencies of current inhalers is 10\% - 30\% for adults and 3\% - 15\% for children. Again \cite{deng2019particle} showed that the deposition of particles in upper, middle and lower lungs varies with breathing rates that vary with various physical activities such as sedentary activities (sitting, lying down etc), moderate activities (walking, cycling etc) and intense activities (running, hiking, swimming etc). Moreover, most of the numerical studies are restricted up to few generations in the upper, middle or lower lung due to the limitation of the computational power and time. On the other hand, experimental deposition studies are restricted to nose, throat, few generations in upper lungs or scaled up version of inner lungs due to its complex and miniaturized geometry. Since the deposition in pulmonary and acinus region decides the response of the drug and its delivery efficiency, it is necessary to estimate deposition fractions in inner lungs experimentally in true scale models and thereafter validate numerical results. 
\par
From an engineering point of view, human lungs is an excellent natural device to study as it generates a wide range of flow conditions starting from $Re = 10^3$ to $Re = 10^{-2}$. This provides an opportunity to study different flows such as turbulent flow in the upper airways, laminar flow in mid airways and creeping flow in lower airways and their effects on particle transport as well as deposition. Thus the necessity of detailed experimental study and the elegance in the physics that is observed in a single human organ motivates and inspires this work.

\section{Objectives}
While the primary purpose of the respiratory system is to enable gas exchange, it also provides a route for pulmonary aerosol transport. The constant variation of airway diameter causes a huge change in $Re$, over five orders of magnitude in every single breath. Since different fluid mechanic conditions give rise to various deposition mechanisms, it is necessary to be able to quantify the total lung deposition along with the regional depositions. The deposition in distal alveolated region seeks attention for quantifying disease propensity as well as therapeutic efficacy. The objective of this proposed work are:

\begin{enumerate}[label=\roman*.]
    \item  To study the effect of bronchiole diameter and aspect ratio, mimicking the lung airways, on aerosol deposition.
    \item  To study the effect of particle Reynolds number ($Re$) on deposition of aerosols.
    \item  To quantify the amount of deposition for different orders of $Re$, starting from $Re \sim \mathcal{O}(10^3)$ (in tracheal region) to $Re \sim \mathcal{O}(10^{-2})$ (in alveolar region).
    \item  Finally, to develop a universal bronchiole scale deposition model, which can form the building block for estimating the total aerosol deposition in lungs. 
\end{enumerate}
    
\section{Overview of the thesis}
The thesis is designed to start with a brief introduction on lung morphology and thereafter discussing the experimental techniques used to quantify the deposition of aerosols in phantom bronchioles. The effect of several parameters like capillary diameter, aspect ratio, Reynolds number ($Re$) etc. on aerosol deposition is discussed in detail in this thesis. Finally, this study ends with developing a universal bronchiole scale deposition model which can estimate the total aerosol deposition in lungs. The chapters in the thesis are organised as follows:

Chapter \ref{chap:lit survey} outlines a survey of the literature on numerical and experimental studies related to deposition of aerosols. The regional and total deposition studies present in the literature (both numerical and experimental) are discussed. The regional deposition studies that are discussed in this chapter contain deposition in mouth/throat, deposition in upper lungs and deposition in lower lungs. Lastly, the gap in the existing literature is stated clearly, indicating the necessity of this study.

{Chapter 3} includes the detailing of the materials used for this experimental study.

\textbf{Chapter 4} discusses the experimental methods used for this study. The first part of this chapter gives a detailed description of the experimental setup. The chapter also discusses the characterisation of the aerosol particles and measurement of aerosol concentration. The end of the chapter highlights the method for analysis of deposition samples.

\textbf{Chapter 5} discusses a presentation of the experimental results and their non-dimensional counter parts. The first part of this chapter deals with the non-dimensionalizing the parameters varied in the experiment, highlighting the uncertainties associated with them. The later part of the thesis discusses the effect of those dimensionless parameters on deposition of aerosols.

\textbf{Chapter 6} proposes a universal bronchiole scale deposition model for estimating total deposition in the lungs which is developed from the experimental data.

The key results obtained from this study is concluded in \textbf{Chapter 7}. This chapter also includes the scope of continuing this work in future based on deposition study in branched capillary instead of straight tubules.

\chapter{Literature Survey}
\label{chap:lit survey}
Aerosol deposition in lungs has attracted the attention of the researchers for several decades, one of the reasons being to understand the complicated nature of aerosol transport in lung airways. Since the lung provides a viable drug delivery route, researchers are keen to predict the regional drug deposition to estimate drug efficacy. Besides this, researchers have also paid attention to the transport of the atmospheric pollutants to regulate health effects. This chapter starts with a brief introduction to the studies carried out to estimate lung morphometry. The later sections of the chapter discuss the existing literature on aerosol deposition in the throat, upper lungs and lower lungs along with the deposition models developed by the researchers.

\section{Determination of lung morphometry}
A preliminary model of lung geometry and dimensions was given by \cite{weibel1963morphometry}, followed by various studies reporting lung morphometry \citep{horsfield1968morphology, phalen1983tracheobronchial,yeh1976tracheobronchial}. These researchers estimated the length, diameter and angle of bifurcation at every generation using silicon rubber casts of dissected mammalian or cadaver human lungs \citep{horsfield1968morphology,yeh1976tracheobronchial, phalen1978application}. The morphometric dimensions of the bronchial tree as reported by \cite{horsfield1971models} is restricted to diameters below $0.7mm$ and no attempt was made to find the mean pathway of the lungs. A detailed measurement study was later reported by \cite{raabe1976tracheobronchial}. This study consists of data pertaining to the length, diameter and bifurcation angles starting from trachea to the terminal bronchioles. But their study is restricted to $10 \% - 25 \%$ of the bronchioles, which makes the acinar structure somewhat incomplete. Morphometry data in the acinar region is very sparse in the literature due to its small dimensions. One of the first acinar measurements was given by \cite{hansen1975human}, which completes the respiratory model and is now widely used.

A realistic lung model with proper dimensions is required to study aerosol deposition, as the bronchial tree model may play a role in determining drug delivery efficiency \citep{mauroy2004optimal}. The above studies assume that the measured fraction of the airways always represent the fraction of missing airways \citep{hofmann2008semi}. Thus, these studies have contributed to the development of a \textit{symmetric deterministic model} or \textit{typical path model}. A few studies reported \textit{asymmetric} or \textit{multi path models}, which were developed by combining the stochastic paths for the missing lower airways with symmetric models \citep{asgharian2001particle, hofmann2008semi}. \cite{hofmann1990monte} used diameter frequency distribution and different correlations between parent and daughter airways to develop multi-path models.
\begin{figure}[hbt!]
\centering
\includegraphics[width=14.5cm,height=10cm]{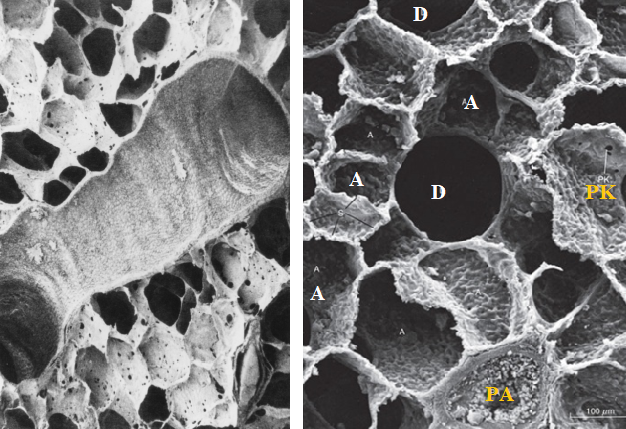}
\caption{Figure showing Scanning Electron Microscope image of a section of human lung parenchyma. A - alveolus; S - alveolar septa; D - alveolar duct; PK - pore of Kohn; PA - small branch of the pulmonary artery. Source: \cite{levitzky2007nonrespiratory,west2012respiratory}}
\label{Figure: Lung SEM}
\end{figure}
Figure \ref{Figure: Lung SEM} shows an electron microscopic image of a section of the human lungs, showing the small bronchioles and alveoli. It gives a good pictorial representation of the structure and dimension of respiratory bronchioles, alveolar ducts and alveolar sacs.

\section{Deposition studies on throat model}

While morphometry has been the focus of a large body of literature, the objective of these studies is to understand breathing and aerosol deposition for therapeutic use. It is necessary for the inhaled drug to reach the pulmonary region of the lungs to show its remedial effects. However, a major fraction of the nebulized drug deposits in the mouth and throat due to its $90^\circ $ bend geometry, which is a matter of great concern. The momentum of aerosol significantly gets affected by mouth and throat geometry which causes significant amount of deposition in that region \citep{longest2008comparison}. The deposition efficiency in the throat region when correlated with initial parameters like droplet size and flow rates, indicates that the inertial impact is the dominant deposition mechanism. Additionally, the inhaler geometry also plays a significant role in deposition in the throat and mouth
\citep{dehaan2001vitro}. An elaborate study was done by the researchers on the deposition of aerosols with an idealized human throat model. The numerical studies of aerosol deposition \citep{li1997aerosol} showed that the deposition in the throat increases with particle diameter. The experimental results for deposition of poly-dispersed aerosol in human throat replica showed good agreement with numerical results generated by k-$\epsilon$ model \citep{stapleton2000suitability}. But the PIV study on molded throat model by \cite{jayaraju2008large} showed good validation with DLS and LES rather than k - $\omega$ RANS model. The deposition for in vivo human throat showed a deposition of $40 \%$ for $1 \mu m$ particles and of $80 \%$ for $10 \mu m$ particles. Effect of Stokes number and flow Reynolds number were examined using gamma scintigraphy and gravimetry for extrathoracic deposition of aerosol using the idealized human mouth and throat replica by \cite{grgic2004regional}. Their study showed the aerosol to be deposited mostly in the larynx and trachea. However, the total deposition was observed to be independent of mouth-throat geometry. It was observed the deposition fraction is largely dependent on inertial parameters, indicating impaction dominated deposition mechanism \citep{grgic2004regional}.
 
\section{Deposition studies in the upper lung}
A significant number of experimental and numerical studies have been focused on aerosol deposition in upper lungs to understand the flow physics and deposition mechanism in airway branching. \cite{deng2019particle} investigated deposition of atmospheric particulates within a one-dimensional numerical trumpet model of lung. They concluded that particles of size $> 2.5 \mu m$ deposit mainly in the tracheobronchial region of the lungs which reduces the pulmonary deposition. The deposition varies heavily with the breathing rate or with the type of activities \citep{deng2019health}. The particle deposition is uniform among different generations for moderate activity, more in lower airways (14 - 16 generations) during sedentary activities and greater in upper airways (3 - 5 generations) during intense activity. This heavy deposition of particulates in tracheobronchial airways can cause different health effects for a person undergoing different activities throughout the day. The photo-metric measurement of inhaled and exhaled iron oxide particles of  of $1 \mu m - 10 \mu m$ by \cite{stahlhofen1980experimental} showed similar result where the deposition is high in the thorax. The study of \cite{bennett2019vitro} included the effect of lung tidal volume and inspiratory to expiratory ratio along with the breathing rate. Their study shows an increase in aerosol deposition with tidal volume, breathing rate and inspiratory time. They concluded that breathing frequency is not the only factor responsible for heavy deposition in the tracheobronchial region. The numerical study of \cite{chakravarty2019aerosol} showed that the increase in breathing frequency enhances the transport rate of aerosol and it is further enhanced when the particle size is reduced.External forces such as gravity and inertia act on inhaled  particles causing non-uniform deposition of aerosols in lung airways. \cite{asgharian2006airflow} The heterogeneous pattern of aerosol deposition in the trachea and bronchus changes with types of nebulizer among which mesh nebulizer is found to be most efficient for drug delivery to the bronchus \citep{xi2018visualization}. 
\par
 A detailed understanding of airflow throughout the lungs is required to locate vulnerable sites of particle deposition and to target specific regions of the lungs with inhaled drugs more efficiently. The transition and the turbulence regime of airflow have a significant effect on localized particle deposition for bifurcating respiratory model. The investigation of \cite{longest2007validating} showed that even though Reynolds number ($Re$) is below the critical limit required for full turbulence, the transitional characteristics and turbulence has a great influence on the flow field and local particle deposition patterns. The experimental determination of aerosol deposition within the lung of human subject or lung model is restricted to the total deposition of aerosol \citep{lizal2015method} for most of the cases although it is not a good indicator of inter-subject variability \citep{devi2016designing}. Researchers have reported in vivo results for inhaling either through mouth or nose for a wide range of particle sizes and flow rates. \citep{heyder1986deposition} The total deposition of aerosols in the lungs increases with an increase in particle size and flow rate \citep{darquenne2016total} which concludes that the deposition in a given region in the lungs is directly proportional to the volume of air delivered to that region \citep{asgharian2006airflow,asgharian2004modeling}. 
 
\section{Deposition study in deep lung model}
The knowledge of aerosol deposition in deep lung is essential in view of advanced strategies of drug targeting, since alveolus provides the communication between the inhaled air and the blood stream \citep{semmler2012nanoparticle,kleinstreuer2008targeted}. A large number of computational studies conducted over past decades provided in-depth understanding of acinar flows and their intrinsic complexities. The acinar flow plays a critical role in influencing the kinematics of inhaled particles. Thus a detailed knowledge of acinar flow phenomena is required in characterizing aerosol deposition in the acinar region \citep{sznitman2013respiratory} as anisotropic wall motion of the acinar airways may influence deposition in deep lung \citep{hofemeier2016role}. The numerical simulation shows enhancement of particle transport rate due to a reduction in particle size, which signifies that diffusion mechanism dominates over advection in the acinar region \citep{chakravarty2019aerosol}. As diffusion is a slow process, the breath holding time plays a significant role in alveolar deposition \citep{khajeh2015deposition}. The numerical study of \cite{koullapis2018efficient} shows particles of size $1 - 5 \mu m$ have the highest possibility of deposition in the acinar region since they remain suspended and sediment during the early stage of exhalation. Thus for effective drug delivery one should give a long pause after inhalation, so that the deposition of the drug increases in the acinar region \citep{koullapis2020towards}.
\par
Most of the experimental studies available in the literature are restricted to the upper generation of lungs. Deposition studies in the deep lung (i.e. G = 20 $\rightarrow$ 23) are rare due to its micro-scale dimensions and complex geometry. The real time visualization of particle dynamics in real lung is also unavailable due to lack of appropriate experimental techniques. With the advent of micro-fluidics \citep{baker2011living,esch2011role,huh2010reconstituting,moraes2012organs}, some insights on acinar deposition are obtained but the results in physiologically realistic acinar environment is still rare \citep{sznitman2013respiratory}. The very small flow rate in acinar region  and the micro size of the airways makes it difficult to mimic the corresponding flow conditions for experimental study. Among the few experimental studies in the existing literature, the experimental results on true scale acinar model by \cite{fishler2015particle} shows a good agreement with numerical data. The bifurcated alveolar ducts that have breathing like wall motion to capture respiratory flows in acinus but it is restricted to constant diameter bifurcations although the variation in diameter have significant effect in deposition. Similarly, \cite{lin2019aerosol} investigated with bifurcated tubules of differing diameter and low $Re$ (between $0.1$ and $1$), but the study is restricted to a few generations of the lower lung that does not completes the acinar deposition model. The convective mixing of the particles found to be maximum near alveolar openings where the streamlines are slow and are prone to deposition \citep{fishler2017streamline}.
\par
While different deposition mechanisms are activated in different regions of the lung owing to differing fluid mechanic conditions, it is important to be able to quantify the total amount of aerosol deposited in the lung \citep{devi2016designing, lizal2015method} as well as deposition by region \citep{ferron1977deposition,stahlhofen1980experimental}. Since inhalation therapy finds application in the treatment of various lung disorders such as asthma, chronic obstructive pulmonary disease (COPD), respiratory infection, lung cancer etc, it is important to quantify the regional deposition, especially in the distal alveolated region to predict therapeutic efficacy. Thus accurate models of regional deposition in the lung are necessary both from clinical and scientific points of view to target specific regions of the lung with greater accuracy \citep{manshadi2019magnetic,dames2007targeted}.  
\par
This work is focused on investigating the parametric effects on aerosol deposition in straight micro-capillaries as well as to develop validated model. The literature pertaining to aerosol deposition in circular pipes is quite mature, but at the same time mostly focused on high $Re$ turbulent flows or laminar flows ($Re \sim 10^1 - 10^2$). Through an experimental study, \cite{kim1984deposition} showed that the aerosol deposition in a straight pipe increases with increasing particle size. However their study is restricted to $\sim 0.5 cm$ tubes with the corresponding $Re$ ranging from $140$ to $2800$. Similarly, the studies of  \citet{montgomery1970aerosol, malet2000deposition, muyshondt1996turbulent, friedlander1957deposition} with higher diameter pipes is associated with very high $Re \sim 10^5$ that addresses deposition in high $Re$ turbulent flows and fails to explain deposition in lung airways. The experiments done by \citet{sehmel1968aerosol} were also in the range of $Re \sim \mathcal{O}(10^{4})$. In all the cases, the deposition was characterized in a range of $Re$ where impaction dominates the deposition process. The lacuna that the current work attempts to fill is to span a wider a range of $Re$ to gain a holistic understanding of aerosol deposition of all the underlying physical processes. The overarching goal is to construct a model that will be able to predict deposition in the entire lung based on extensive measurements of aerosol deposition in phantom bronchioles spanning five order of magnitude of Reynolds numbers. The study also points to a minimal set of dimensionless variables that are sufficient to capture the deposition efficiency over wide range of $Re$ ($10^{-2} \leqslant Re \leqslant 10^{3}$) that defines the novelty of this work.
\chapter{Experimental Methods}
 This chapter highlights the experimental setup and the procedure of collecting and analysing deposition samples in details. The experiments are carried out with capillaries of varying diameters ($0.3mm - 2mm$) mimicking $7^{th}$ to $23 ^{rd}$ generation of human lungs. The experimental $Re$ ranges from $Re$ ($10^{-2} \leqslant Re \leqslant 10^{3}$) that mimics the flow conditions of the entire lungs for a single breath.
 
\section{Experimental setup}

\begin{figure}[h!]
\centering
\includegraphics[width=14cm,height=8.2cm]{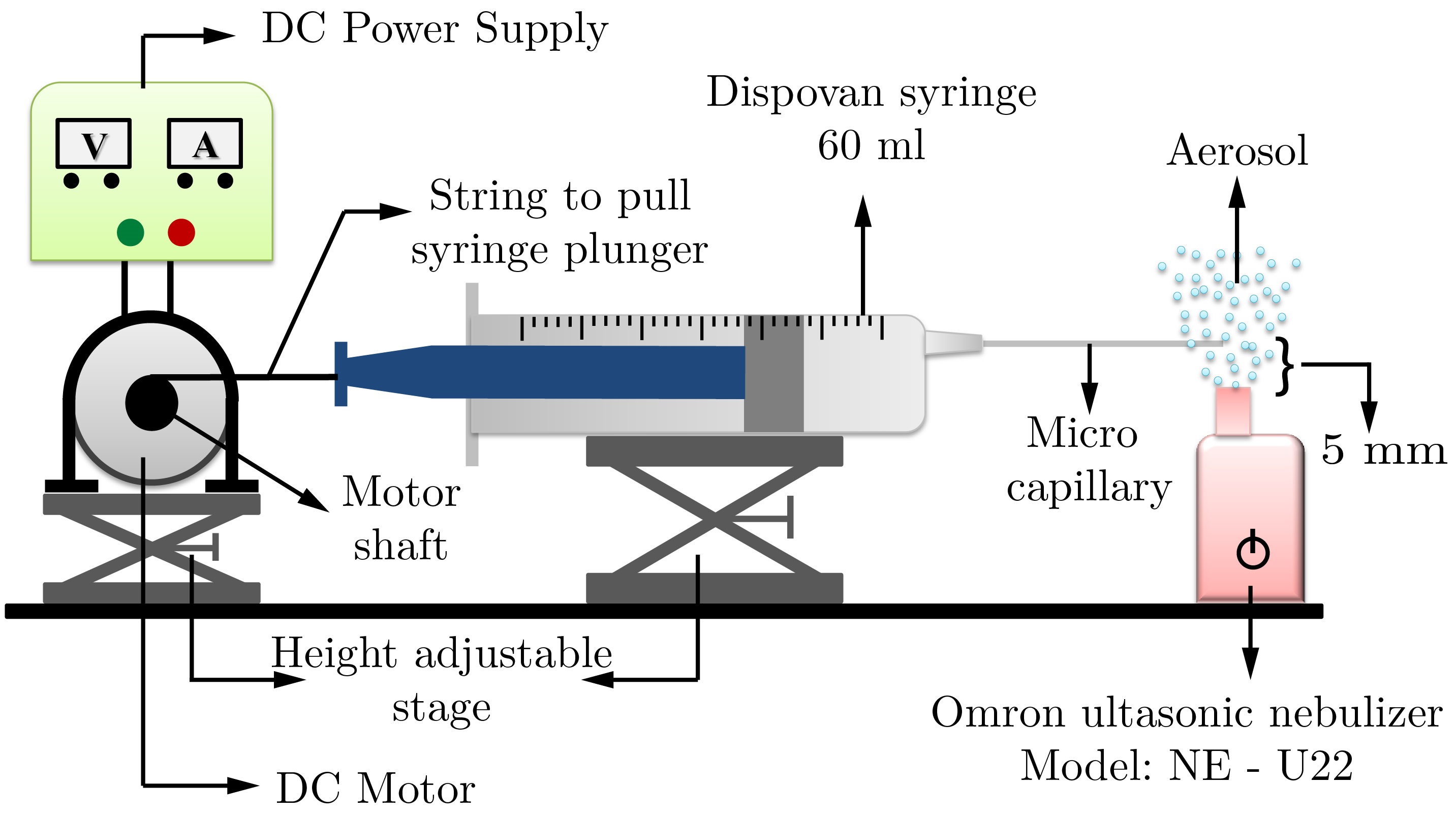}
\caption{Schematic of the experimental setup consisting of aerosol generator, micro-capillary (phantom bronchiole), syringe with a motor arrangement to draw the plunger at a constant velocity. The inlet of the bronchiole is placed at a height of $5 mm$ above the geometric center of the nebulizer exit. Different flow conditions through the micro-capillaries were achieved by controlling the rotational speed of the $12 V$ DC motor using a $32 V$, $2 A$ DC power supply.}
\label{Figure: Expt Schematic}
\end{figure}

The experimental setup shown in Figure \ref{Figure: Expt Schematic} consists of a horizontal micro-tubule (phantom bronchiole) which is attached to a syringe. Different diameters and lengths of these tubules are intended to mimic different generations of bronchioles in the lung as indicated in Table \ref{tab:Capillary diameter}.
\begin{table}[ht!]
\centering
\caption{\label{tab:Capillary diameter} A table listing bronchiole diameter as a function of the branching generation in the lung. Typical $Re$ is calculated based on tidal volume of $500 ml$ and a $4 s$ breathing cycle.}
\begin{tabular}{ccc}
\hline \hline
Typical bronchiole dia (mm) & Lung generation & Typical $Re$ \\[0.5ex]
\hline 
10 & 0 $\longrightarrow$ 6 & $10^3 - 10^2$\\[0.5ex]
1.5 & 7 $\longrightarrow$ 10 & $40$\\[0.5ex]
1 & 11 $\longrightarrow$ 13 & $5$ \\[0.5ex]
0.5 & 14 $\longrightarrow$ 21 & $ 10^{-1} - 10^{-2}$\\[0.5ex]
0.3  & 22 $\longrightarrow$ 23 & $7 \times 10^{-3}$\\[0.5ex]
\hline \hline
\end{tabular}
\end{table}
Polytetrafluoroethylene (PTFE) tubules of different diameters: $0.3 mm$, $0.5 mm$, $1 mm$, $1.5 mm$ and $2 mm$ and with good surface finish are purchased from Cole-Parmer\textsuperscript{\textregistered}, were used as phantom bronchioles. These tubules were translucent, chemically inert, have low water absorption property $< 0.01\%$ and have very low coefficients of friction ($\sim 0.05 - 0.1$) which is third lowest of any solid . The non-sticky nature of this material helps in complete removal of the deposited aerosol by washing with water.

A syringe plunger intended to drive the flow is connected to the shaft of a DC motor. A DC motor of ratings $10 rpm$ at $12 V$ and $30 rpm$ at $12 V$ is used to achieve different suction rates. The flow rates has been verified several times before experiment by tracing the location of plunger within stipulated time. With the rotation of the motor shaft, the plunger is actuated at a constant rate to draw the aerosol exiting the nebulizer into the PTFE tubule. The rotational speed of the motor is varied to obtain different flow rates. The rotational speed is in turn, varied by varying the voltage at the motor terminals with the help of a $32 V$, $2 A$, DC power supply. For very low suction rates ($Re \ll 1$), a syringe pump is used. Interestingly, the lowest $Re$ experiment involved an ultra low flow rate which required that a single experiment be performed over a duration of $5 hours$ for reliable measurements. An ultrasonic mesh type nebulizer (Omron Model: NE-U22) is used to generate a finely atomized, gently rising aerosol plume. The setup is mounted on a height adjustable stage to ensure proper motion of the plunger and to maintain the micro-tubule at a height of $5 mm$ from the nebulizer exit, where all drop size measurements were performed.

\section{Preparation of aerosol liquid}
The aerosol liquid composed of water as the fluid medium and was loaded with boron quantum dots \citep{reed1989quantum} , which acted as fluorophores. The photo-physical properties of quantum dots, including carbon quantum dots (CDs), depend on their size \citep{song2016invisible} as its band gap originates from quantum confinement. The tuning of band gap, which is measured by quantum yield (QY), can be achieved by incorporating trap sites while introducing functional groups during the synthesis of CDs. In order to increase QY, trap sites of CDs are often doped with hetero atoms such as nitrogen \citep{mi2013nitrogen}, boron \citep{fan2013electronic} or phosphorous \citep{ananthanarayanan2014facile} depending on the application.

\begin{figure}[ht!]
\centering
\subfigure[]{\includegraphics[width=5.3cm,height=8cm]{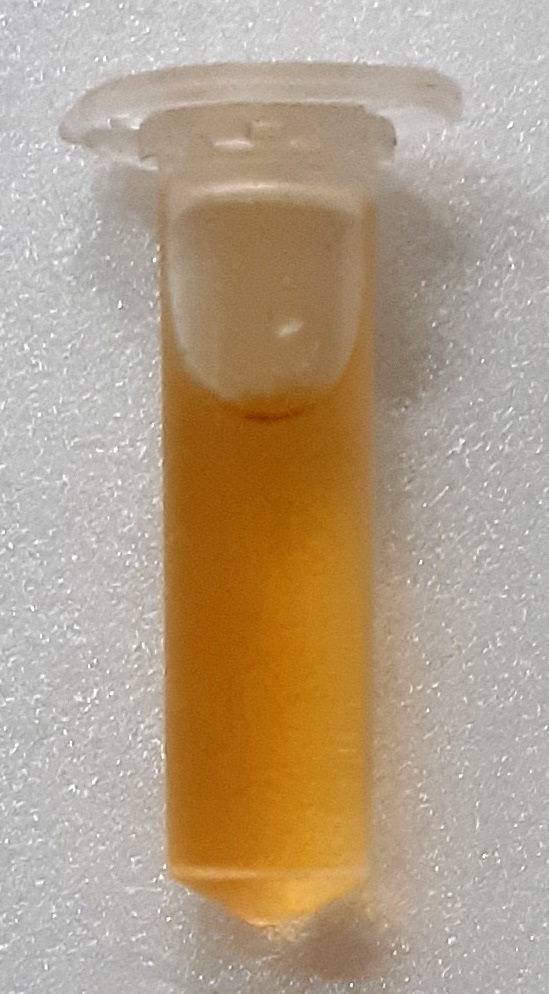}}
\subfigure[]{\includegraphics[width=6.2cm,height=8cm]{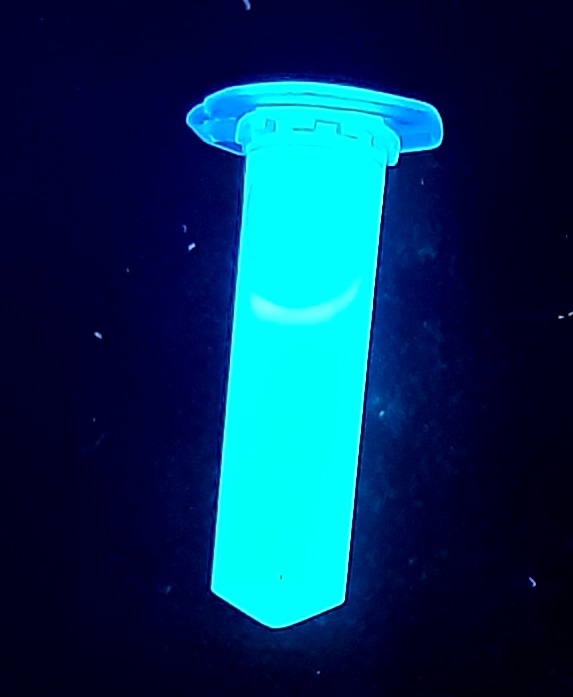}}
\caption{Images of vials (a) containing a solution with freshly prepared Boron doped Carbon Dots (BCD) and (b) showing emission of blue light by BCD when excited by ultraviolet light of wavelength $350 nm$.}
\label{Figure: BCD}
\end{figure}
\begin{figure}[ht!]
\centering
\subfigure[]{
\includegraphics[width=10cm,height=7.5cm]{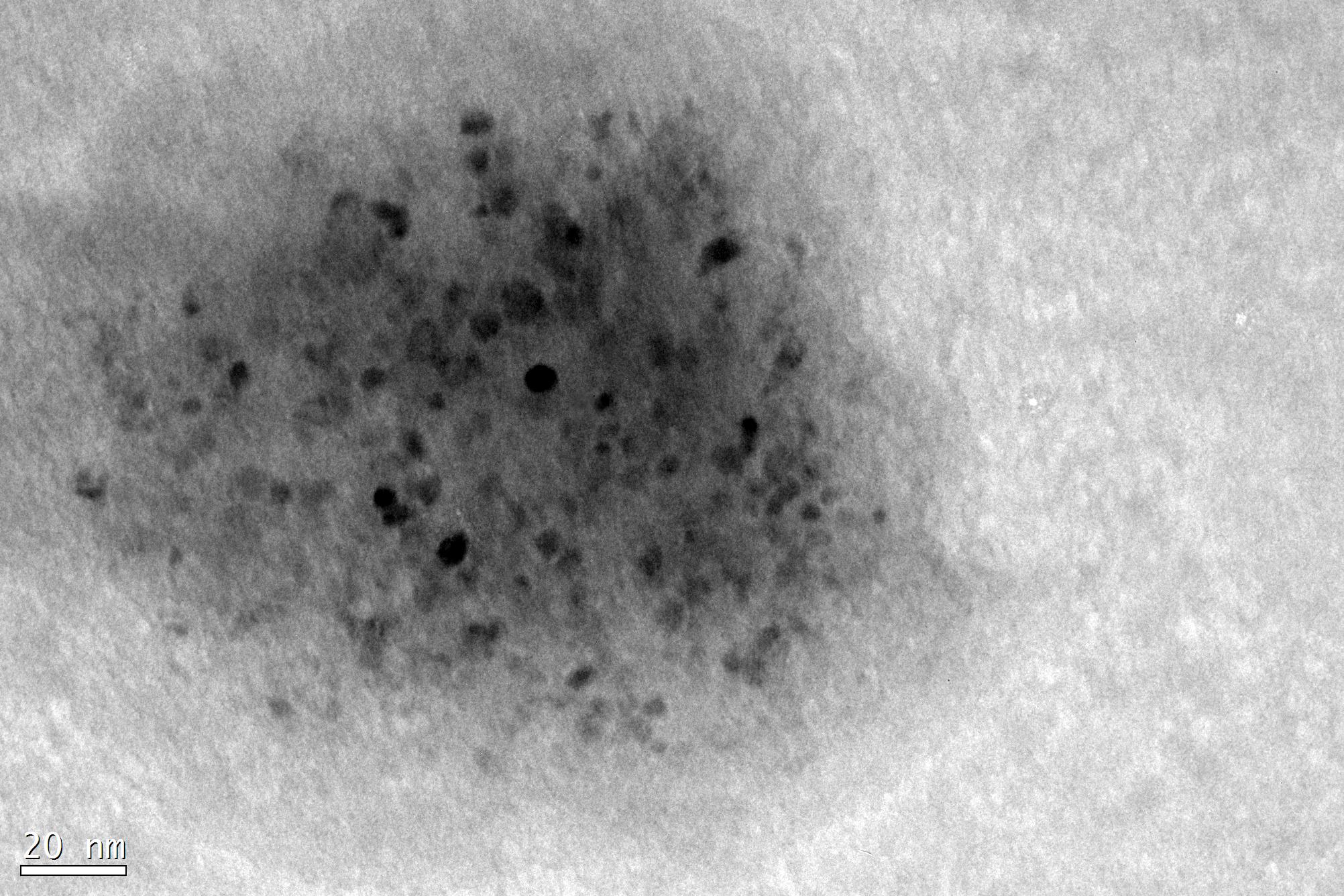}}
\subfigure[]{
\includegraphics[width=10cm,height=7.5cm]{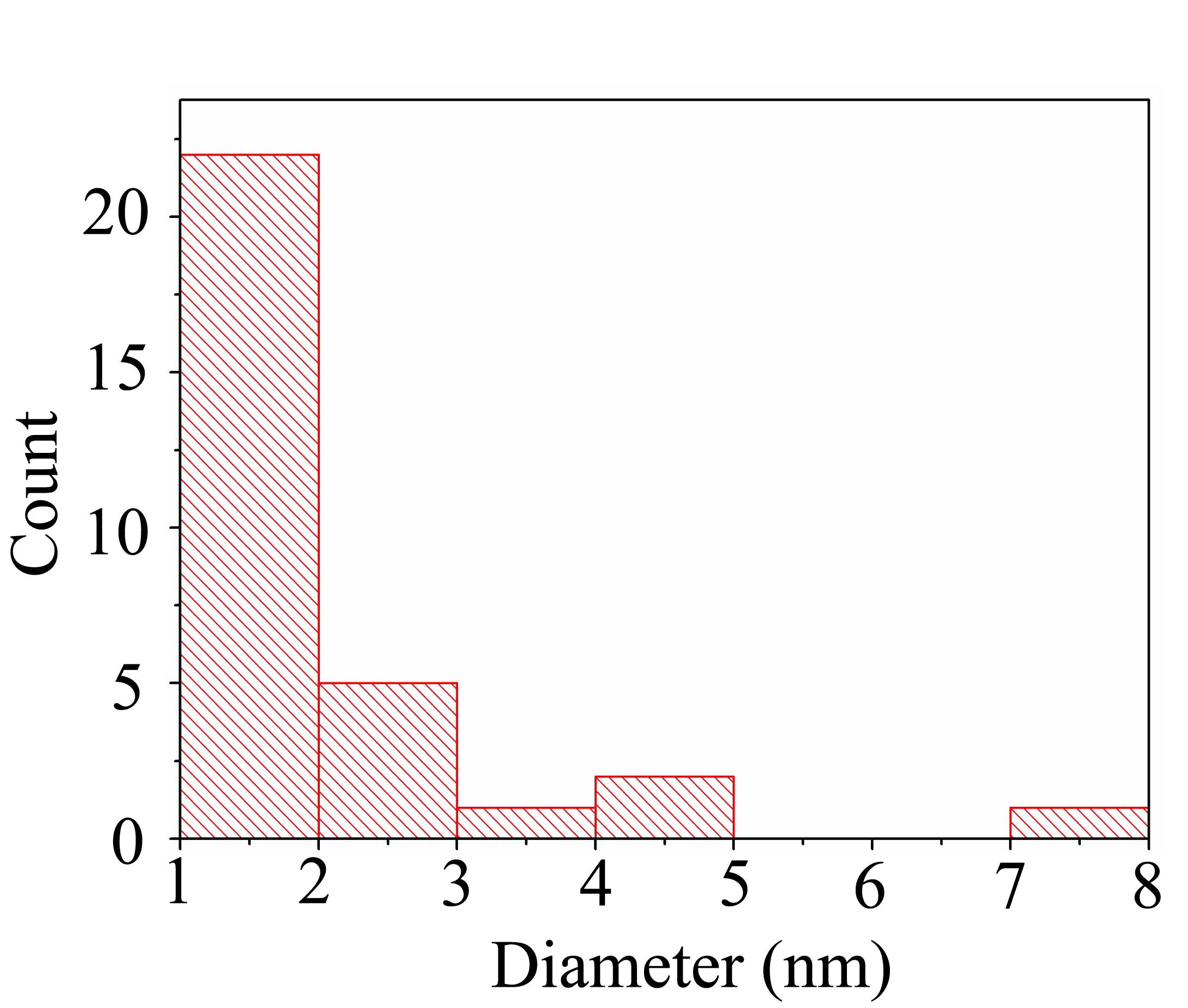}}
\caption{(a) Transmission Electron Microscopy (TEM) image of Boron doped Carbon Quantum Dots. Scale is indicated as a white bar. (b) Histogram of the size distribution of BCD estimated from the TEM images. The BCD distribution is nearly mono-disperse.}
\label{Figure: TEM}
\end{figure}
\par
The synthesis of Boron doped Carbon Quantum Dots (BCDs) was done in the laboratory through a bottom up-process. Care was taken to ensure that the Deionised (DI) water used for synthesis of CDs and BCDs has a pH of 7. All the chemicals used for synthesis were sourced from Sigma-Aldrich \textsuperscript{\textregistered}, Merck\textsuperscript{\textregistered}, Fischer-Scientific\textsuperscript{\textregistered}and Spectrochem\textsuperscript{\textregistered}. Equimolar concentration solutions at $0.5 M$ each of boric acid (Boron precursor) and glucose (carbon precursor) were prepared with $10 ml$ of DI water and mixed using a magnetic stirrer at $300 rpm$ for 15 minutes. The solution was transferred to a glass bowl and treated under commercial microwave (IFB) radiation of $700 W$ for $5 min$. The resulting solid was then dried in vacuum to remove all volatiles and dispersed in $500 ml$ of DI water to form an olive green solution as shown in Figure \ref{Figure: BCD}(a). The larger particles were sorted out by centrifugation at $1500 g$ for $15 min$ and filtered in vacuum using a $10 kDa$ filter. The resultant solution emits blue light when excited by a wavelength of $350 nm$ as shown in Figure \ref{Figure: BCD}(b). The morphological characteristics of BCDs were investigated by Transmission Electron Microscopy (TEM) (see Figure \ref{Figure: TEM}(b) for a representative image). Their size distribution was estimated from the TEM images with the help of ImageJ \textsuperscript{\textregistered}. Figure \ref{Figure: TEM}(b) shows the obtained size distribution. As can be seen, the BCDs are nearly mono-disperse and range in size from $1 nm$ to $2 nm$ (Figure \ref{Figure: TEM}b) which is approximately 5 orders smaller than the smallest dimension of the tubule used in the experiment. Since the BCDs were nearly mono-disperse, the emitted fluorescence spectrum is likely to be in a narrow wavelength band.

\section{Generation and characterization of aerosol plume}
\begin{figure}[hbt!]
\centering
{\includegraphics[width=12cm,height=10cm]{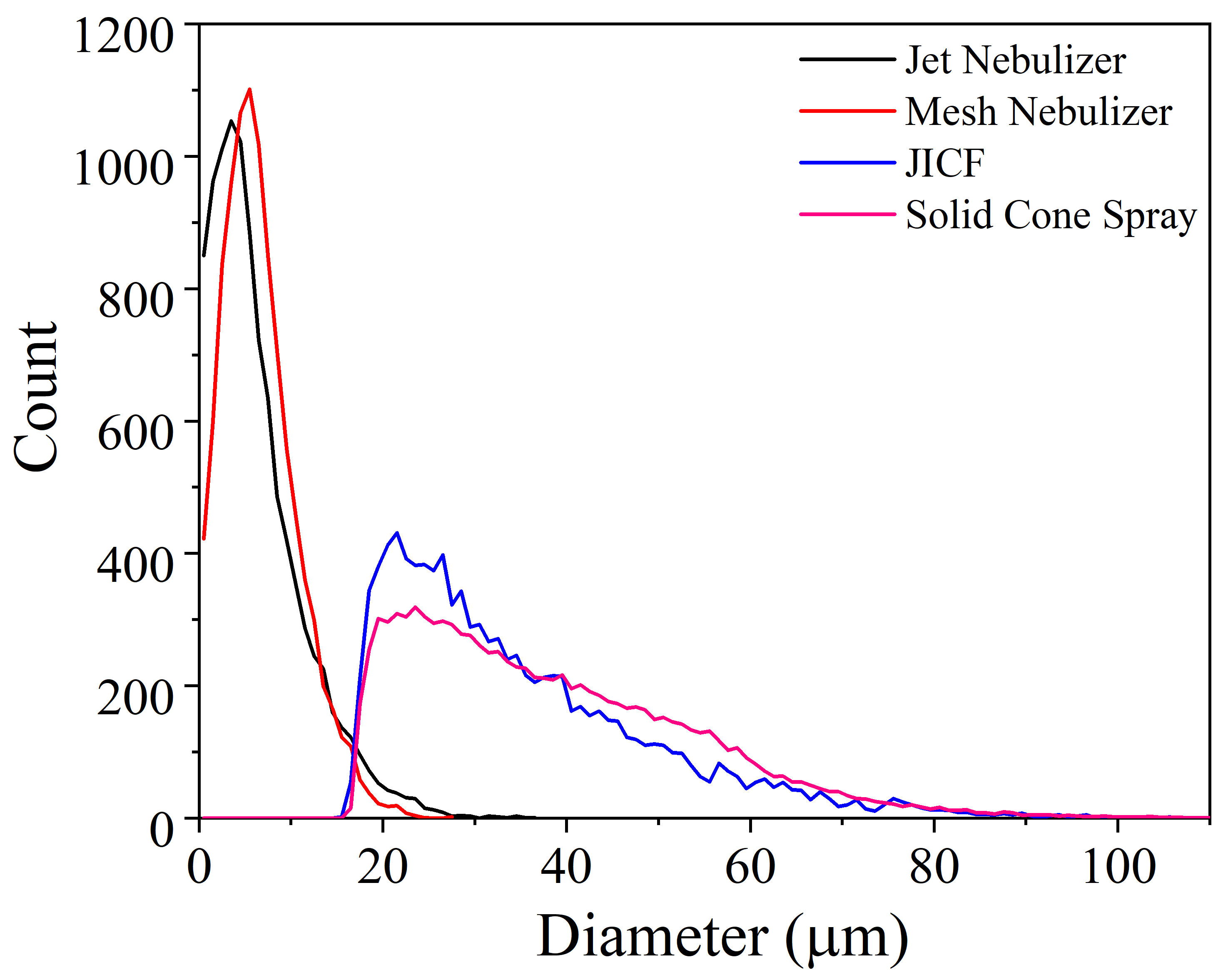}}

\caption{Plot of droplet size distribution for Omron\textsuperscript{\textregistered} make jet nebulizer, mesh nebulizer, JICF \citep{mallik2020phase} and solid cone spray (Make: Delevan\textsuperscript{\textregistered}). In the figure it is evident that the size distribution of the spray and JICF spans a wide range of diameters where as the distribution of the nebulizers is comparatively mono dispersed. }
\label{Figure: Different atomizers}
\end{figure}

There are several methods of droplet generation from a bulk liquid according to the existing literature. The disintegration of bulk liquid requires external forces that contributes in overcoming the surface energy and provide kinetic energy to the droplets. The generation of droplets using pressure swirl or air-blast atomizers and shear break up of jets by air is common in different industrial applications. The use of ultrasonic mesh and compressed air for atomization is common in fogger, humidifier and drug nebulizers. In case of conventional atomizers and breakup of jet in cross flow (JICF) the breakup processes under goes ligament formation, primary atomization and secondary atomization \citep{mallik2020phase}. The ligament formation zone consists of big blobs of liquids disintegrated from the bulk liquid, which further breakup to form droplets. This is called primary atomization. The primarily atomized drop undergoes further breakup to form smaller droplets that are stable due to higher surface energy. This different stages of breakup for conventional atomizers and JICF gives rise to a wide range of droplet sizes. On the contrary the ultrasonic or compressed air atomization do not under goes different stages of breakup which results in generation of mono dispersed drops. The figure \ref{Figure: Different atomizers} shows a comparative study of droplet size distribution using different atomization process. The size distribution of Omron\textsuperscript{\textregistered} make jet and ultrasonic mesh nebulizer, JICF \citep{mallik2020phase} and solid cone spray (Make: Delevan\textsuperscript{\textregistered}) is estimated using Phase Doppler Particle Analyzer (PDPA). It is evident from the figure that nebulizers produces mono dispersed drops of smaller sizes in comparison to the other conventional atomization techniques.
\par
In this study, a mesh type ultrasonic nebulizer was used to generate the aerosol plume at a constant rate of $0.25 ml/min$. The droplet size and velocity distributions of the aerosol plume generated by the nebulizer were characterized using a TSI\textsuperscript{\textregistered} make PDPA. It is a non-intrusive, laser-based, single particle and point measurement system which works on the principle of interferometric particle sizing. The optical settings employed for the PDPA are given in the table \ref{tab:PDPA settings}. Since accurate size measurement depends on phase difference of photo detectors, phase calibration was periodically performed to avoid unexpected phase delay. The optical settings of the PDPA were adjusted such that the particle diameter measurement range is $0.5 \mu m$ to $165 \mu m$ with an estimated accuracy of $\pm 0.1 \mu m$ over the entire range. A wide range of velocity measurements from $-100 m/sec$ to $200 m/sec$ was also possible through the appropriate band pass filter choice. As a result, drop size and velocity were measured with an accuracy of $\pm0.2 \%$. Finally, the photo multiplier tube (PMT) voltage is chosen such that it does not add noise to the data while producing a good data rate.
\begin{table}[ht!]
\caption{\label{tab:PDPA settings} Optical settings of PDPA}
\centering
\begin{tabular}{lcr}
\hline \hline
Optical settings & Values \\
\hline
Transmitter Wavelength & $532 nm$  \\ 
Transmitter Focal Length  & $363 mm$ \\
Laser Beam Separation & $50 mm$  \\
Laser Beam Diameter & $2.10 mm$  \\
Beam Expander Ratio & 1 \\
Beam Waist & $117.09 \mu m$ \\
Fringe Spacing & $3.8715 \mu m$ \\ 
Bragg Cell Frequency & $40 MHz$\\
Off - axis angle & $43 ^\circ$\\
Mode of scattering & Refraction\\
Refraction index & $1.33$\\ [0.5ex]
\hline \hline
\end{tabular}
\end{table}
\par
\begin{figure}[h!]
\centering
\includegraphics[width=10cm,height=9cm]{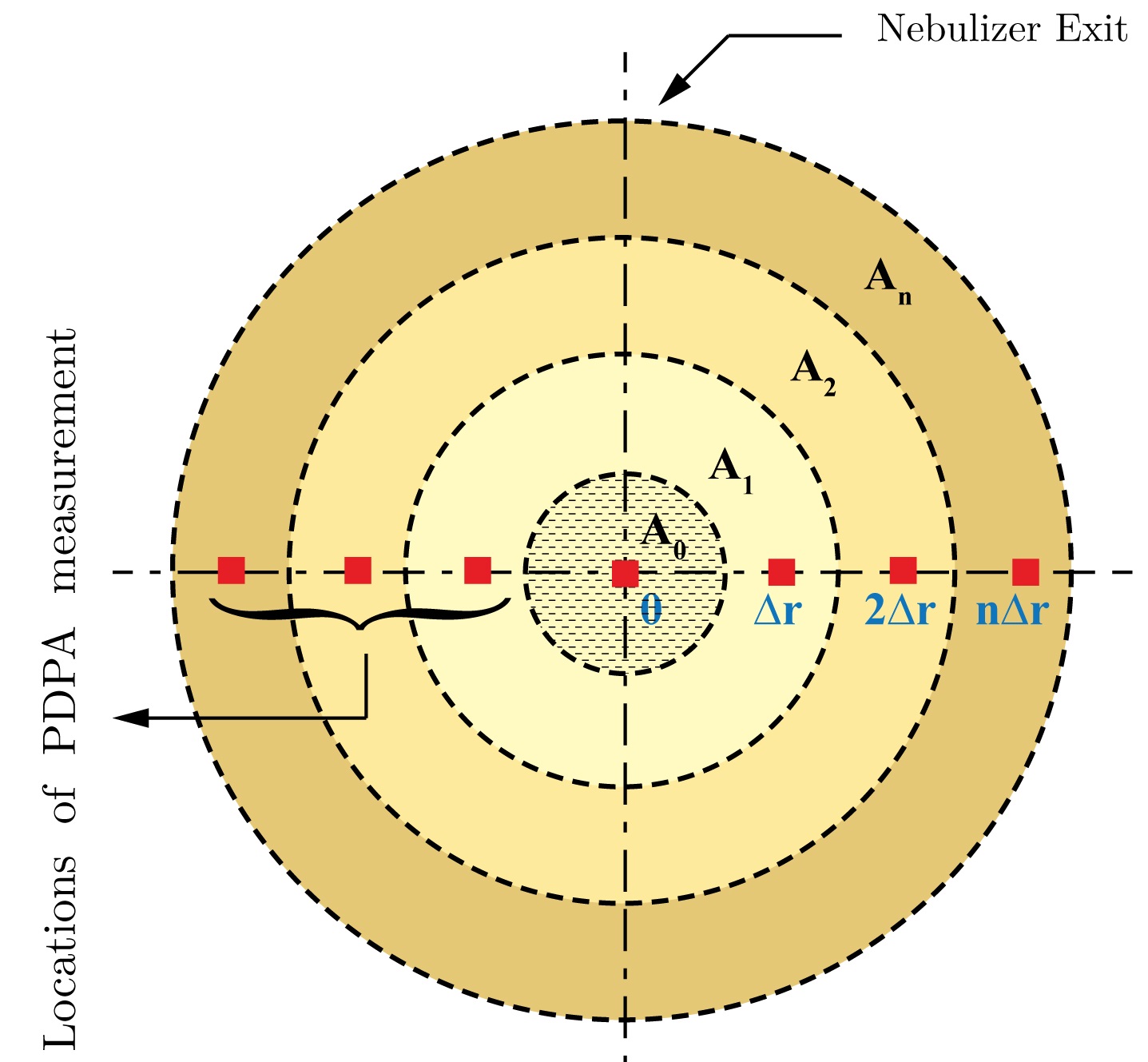}
\caption{Schematic of measurement location of drop size and velocity, $5mm$ above the nebulizer exit, using Phase Doppler Particle Analyzer}
\label{Figure: PDPA Measurement Loc}
\end{figure}

The diameter and velocity of the aerosol exiting the nebulizer is measured at different radial locations, $2 mm$ apart, at an axial distance of $5 mm$ from the nebulizer exit as shown in figure \ref{Figure: PDPA Measurement Loc}. 
\begin{figure}[ht!]
\centering
\subfigure[]{\includegraphics[width=10cm,height=7.5cm]{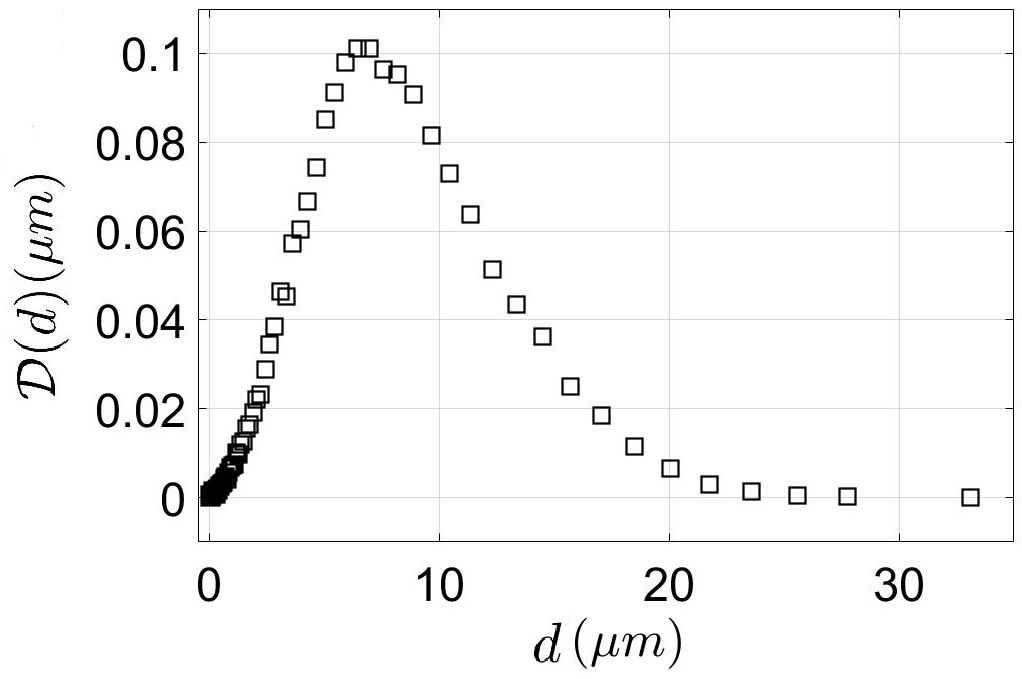}}
\subfigure[]{\includegraphics[width=10cm,height=7.5cm]{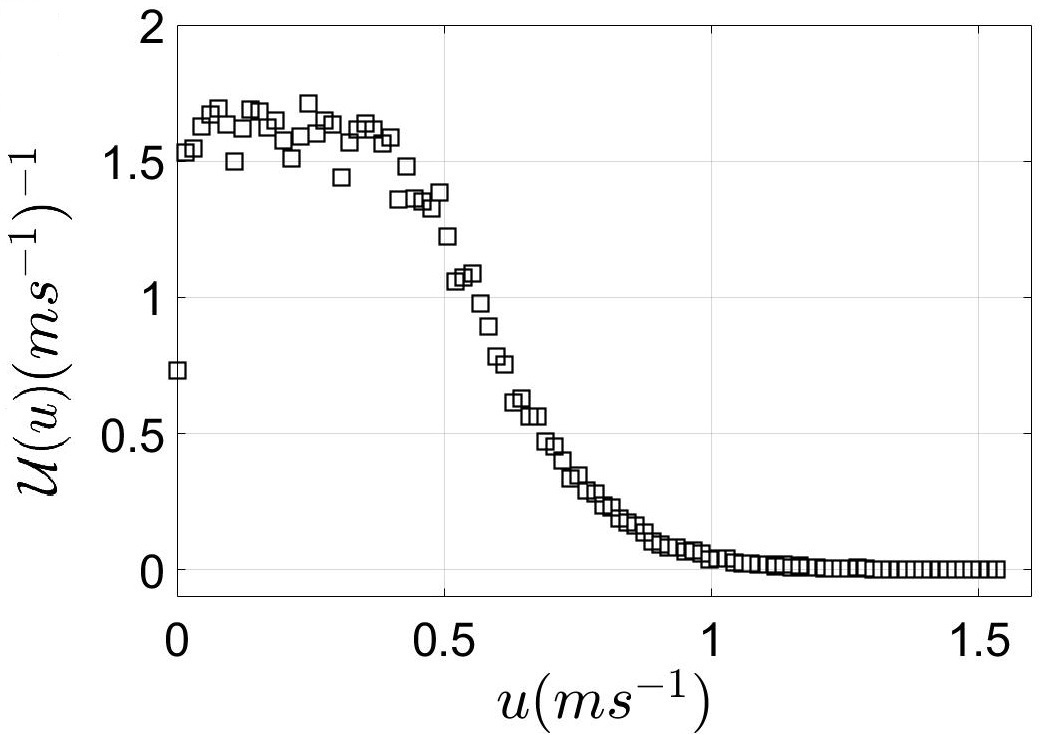}}
\caption{Plot of (a) the global diameter probability distribution function and (b) the global velocity probability distribution function, both measured using PDPA. The most probable drop size of size $6.5 \mu m$. Most drops were moving at a velocity less than $0.5 m s^{-1}$ indicating that the aerosol plume was finely atomized and gently rising, as desired.}
\label{Figure: Global Velocity}
\end{figure}
For ensuring high statistical reliability of the PDPA measurement, 10000 drops were sampled at each measurement location. PDPA only yields point-wise drop size distribution data. A global size and velocity distribution, characteristic of the entire nebulizer was calculated from the point-wise data following the method of Tratnig et al. \cite{tratnig2010drop} and Dhivyaraja et al. \cite{dhivyaraja2019dynamical}. The global diameter $pdf$, $\mathcal{D}(d)$ and velocity p.d.f, $\mathcal{U}(u)$ is given by:
\begin{equation}
   \mathcal{D}(d) = \frac{\sum\limits_{1}^{j} \dot{n}(r_j,D_i) 2\pi r_j \Delta r}{\sum\limits_{1}^{j}\sum\limits_{1}^{i} \dot{n}(r_j,D_i) 2\pi r_j \Delta r},
 \end{equation}
 \begin{equation}
  \mathcal{U}(u) = \frac{\sum\limits_{1}^{j} \dot{n}(r_j,u_i) 2\pi r_j \Delta r}{\sum\limits_{1}^{j}\sum\limits_{1}^{i} \dot{n}(u_j,D_i) 2\pi r_j \Delta r}.
\end{equation}
\par
 Here, $D_i$ is the aerosol size and $u_i$ is the axial component of the aerosol velocity and g is calculated based on number flux density of the drops for diameter and velocity given by $\dot{n}(r_j,D_j)$ and $\dot{n}(r_j,u_j)$ respectively.
The global probability density functions of both drop size and velocity are true representations of nebulizer performance since they are insensitive to external factors. The global aerosol drop size $pdf$ ($\mathcal{D}(d)$) is shown in Figure \ref{Figure: Global Velocity}(a). It shows a maximum probability for $6.5 \mu m$ drops which is taken as the characteristic droplet size in the aerosol plume. The velocity $pdf$ ($\mathcal{U}(u)$) in figure \ref{Figure: Global Velocity}(b) denotes the most probable velocity occurs for $0 < u (m/s) < 0.5$. The mean velocity from this $pdf$ was found to be $0.44 m/s$. From these measurements one can conclude that a finely atomized, gently rising aerosol plume was formed by the nebulizer. 

\par
\begin{figure}[hbt!]
\centering
{\includegraphics[width=14cm,height=11cm]{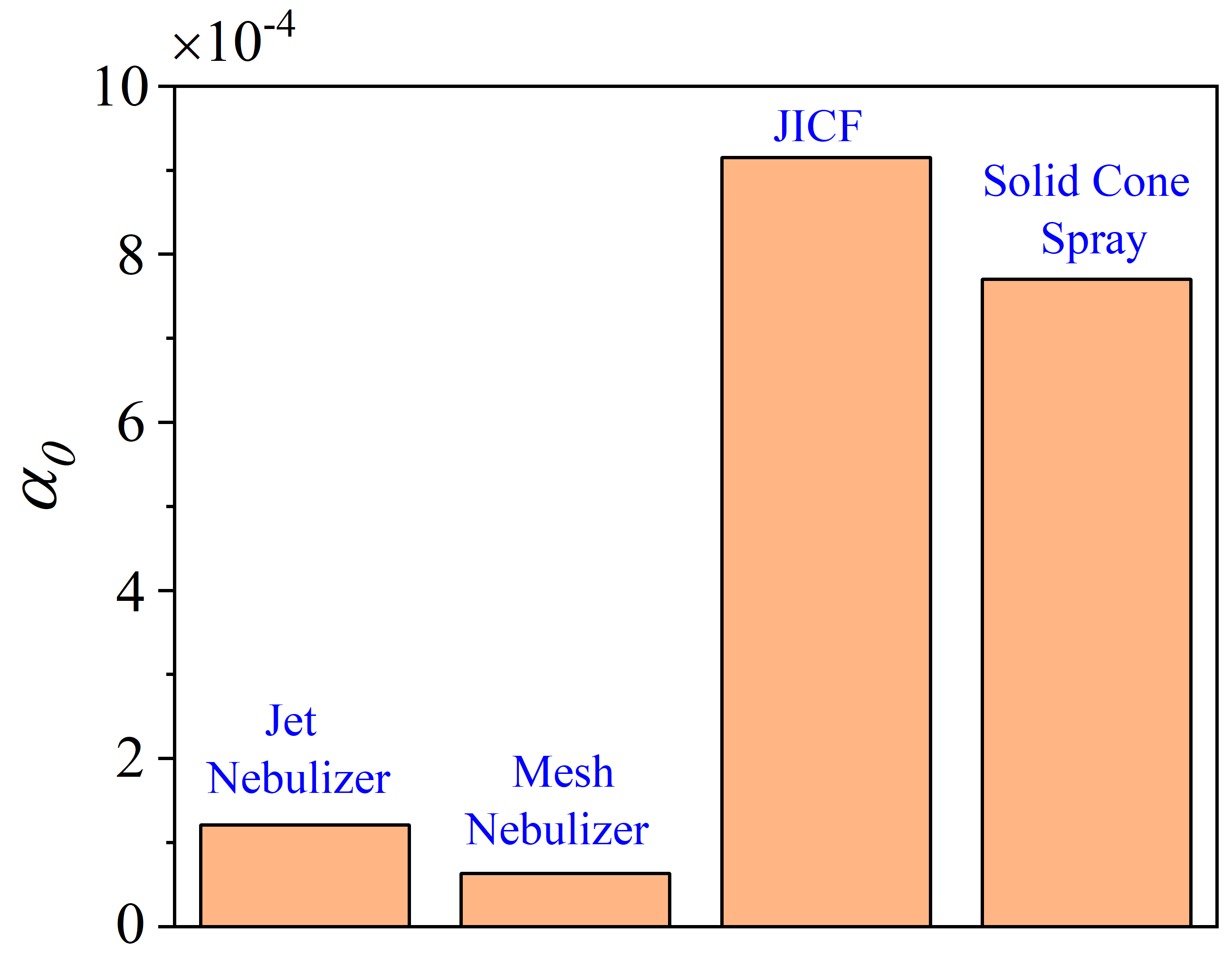}}

\caption{Bar plot showing concentration ($\alpha_0$) of the generated drop for jet nebulizer, mesh nebulizer, JICF and solid cone spray. Among all, mesh nebulizer is found to have lowest droplet concentration of $6.26 \times 10^{-5}$ that confirms the aerosol plume to be dilute and non - interacting.}
\label{Figure: Different atomizers concentration}
\end{figure}

The deposition fraction is measured in each experiment as the ratio of the deposited aerosol to the amount of aerosol that is present in the volume of air drawn in to the bronchiole. The aerosol volume fraction ($\alpha_0$) from the nebulizer at a point is estimated directly from the PDPA measurements. The figure \ref{Figure: Different atomizers concentration} shows that the concentration ($\alpha_0$) of mesh nebulizer is least, $6.2 \times 10^{-5}$, among the other atomizers. Since the $\alpha_0$ is $\sim \mathcal{O} (10^{-5})$, the aerosol plume can be construed to be extremely dilute and consisting of non-interacting droplets.

\section{Preparation and analysis of deposition samples}

The fluorescence signal from the BCDs in the sample was measured using a Horiba FluoroMax\textsuperscript{\textregistered} spectrofluorometer. The instrument consists of a $150 W$ Xenon arc lamp, which were self-calibrated for all wavelength drives and slits. The fluorescence detector consisted of a Photo Multiplier Tube (PMT) which can capture emission wavelengths from $185 nm$ to $850 nm$ with an accuracy of $\pm 0.5 nm$ and a repeatability of $0.1 nm$. The water Raman signal to noise ratio was found to be $6000:1$, as calculated by the First Standard Deviation (FSD) method, and $16000:1$, as calculated by Root Mean Square (RMS) method. This ensured that the fluorescence signal from the BCDs was not confounded by signal from other sources.
\par
The experiment involved an aerosol plume flowing through the PTFE micro-capillaries at a prescribed flow rate. The deposited aerosols within the capillary were then flushed thoroughly with $5 ml$ of DI water to prepare the samples for measuring aerosol concentration. From this, a sample volume of $2 ml$ of the solution is taken in the cuvette and excited with the wavelength of $340 nm$ with a slit width of $3 nm$. The emission intensity was recorded for different wavelengths varying from $405$ to $610 nm$, with a slit width of $5 nm$. The sensitivity of the instrument was checked for an empty cuvette, a cuvette containing DI water as well as for a cuvette containing fluorescence samples for the lowest deposition recorded. It was found that the intensity of the water containing fluorescence sample is at least one order higher than that for normal DI water. The intensity obtained for various aerosol deposition measurements is compared with the intensity obtained from $1 ml$ of aerosol liquid (i.e. BCD) dissolved in the same volume of DI water used for flushing the capillaries. This is taken as the reference value for estimating the deposition in the tubule. Finally, the concentration of the deposited aerosol was calculated from the fluorescence measurement. 
\chapter{Results and Discussion}

This chapter describes the results obtained from the experimental study of aerosol deposition in phantom distal bronchioles. The first part of this chapter introduces the deposition parameters considered in the experiment and their dimensionless forms. The latter part contains an in-depth discussion of the obtained results, describing the effect of individual parameters on deposition.

\section{Experimental Parameters}
The deposition of nebulized particles of mean diameter $6.5 \mu m$ is investigated for micro-capillaries of differing lengths and diameters to mimic different generation bronchioles. The aerosol properties such as kinematic viscosity, droplet size, concentration etc. \citep{ferron1994aerosol} and the tubules dimensions greatly affects the deposition. The deposition concentration of aerosol in these phantom bronchioles can be expressed as a function of several parameters as

\begin{equation}
    d = f (L,D,Q,T,\nu,D_{10},\alpha_0).
\label{eqn: Deposition parameters}
\end{equation}
Here,\\
$d$ = the measured aerosol deposition ($ml$),\\
$L$ = the length of bronchiole ($mm$),\\
$D$ = the bronchiole diameter ($mm$),\\
$Q$ = the volume flow rate ($ml/s$),\\
$T$ = the time duration of the flow ($s$),\\
$\nu$ = the kinematic viscosity of air ($m^2/s$),\\
$D_{10}$ = the mean droplet diameter ($\mu m$).\\
$\alpha_0$ =is the volume fraction of aerosol from nebulizer ($ml$ of aerosol per $ml$ of space),
\par
From a careful dimensionless analysis, one can identify the following relevant dimensionless parameters from the above parameters. These dimensionless parameters are defined in Table \ref{tab:dimgroups}.

\begin{table} [hbt]
\caption{\label{tab:dimgroups} Definition of dimensionless parameters}
\centering
\begin{tabular}{lcr}
\hline \hline
Dimensionless parameter & Definition \\
\hline
Deposition fraction ($D_F$) & $d/(\alpha_0 Q T)$  \\
Aspect Ratio ($\overline{L}$) & $L/D$  \\ 
Dimensionless bronchiole diameter ($\overline{D}$)  & $D/D_{10}$ \\
Reynolds number ($Re$) & $4Q \big/ (\pi\nu D)$  \\
Dimensionless Time ($\overline{T}$) & $4QT \big/ (\pi D^3)$  \\[0.5ex]
\hline \hline
\end{tabular}
\end{table}

\par
Aerosol deposition has been investigated for a wide range of flow conditions, $10^{-2} \leqslant Re \leqslant 10^3$, mimicking the flow in different generation bronchi and bronchioles.  The total volume of air that has passed through the bronchiole during the experiment is $QT$, which is the flow rate multiplied by the breathing time. As a result the total volume of aerosol that entered the bronchiole is $\alpha_0QT$. The deposition fraction ($D_F$) defined as the fraction of the exposed aerosol ($\alpha_0QT$) that has been deposited in the bronchiole, is likely to increase with increasing length of the bronchiole and time of exposure \citep{asgharian2004modeling,koullapis2020towards,khajeh2015deposition}. Thus to normalize for these effects, a dimensionless deposition fraction per unit (dimensionless) length and (dimensionless) time, $\delta$ is defined as \begin{equation}
   \delta = \frac{D_F}{\overline{L}  \overline{T}}
   \label{eqn: delta}
\end{equation}

\noindent Equation \ref{eqn: Deposition parameters} can be rewritten in terms of the dimensionless parameters in Table \ref{tab:dimgroups} as
\begin{equation}
    \delta=G(Re,\overline{D},\overline{L},\overline{T},\alpha_0)
\label{eqn:dimrel}
\end{equation}
It is an endeavour of this work to identify a physics-consistent and universal function $G$ as in equation \ref{eqn:dimrel} from experimental data.

The experiments are repeated several times (minimum 5 times, in few cases 8 - 10 times)  for different $Re$ and aspect ratio to ascertain repeatability of deposition fraction in the phantom bronchioles. The maximum estimated uncertainty  for all the measured parameters was within $\pm 5\%$ calculated based on \citep{kline1953describing,moffat1988describing} formula as  given in equation \ref{uncertanity}.
 \begin{equation}
 \label{uncertanity}
\omega_{R} =\bigg [\bigg(\frac{\partial R}{\partial x_1}{\omega_1}\bigg)^2 + \bigg(\frac{\partial R}{\partial x_2}{\omega_2}\bigg)^2 + \dots + \bigg(\frac{\partial R}{\partial x_n}{\omega_n}\bigg)^2\bigg]^\frac{1}{2}
\end{equation}
 Here $\omega_R$ is the uncertainty in the result and $\omega_1$, $\omega_2$, \dots $\omega_n$ are uncertainties in the measurement of independent variables. R is given by $R = R(x_1, x_2, x_3, \dots x_n)$ where $x_1$, $x_2$, \dots $x_n$ are the independent variables. The values are given in the table \ref{tab:uncertainity}. The effect of the various dimensionless groups on the dimensionless deposition, $D_F$ and $\delta$ are discussed in the later section.

\begin{table} [ht!]
\caption{\label{tab:uncertainity} Uncertainty in the measured and calculated parameters.}
\centering
\begin{tabular}{lcr}
\hline \hline
Derived parameters & Estimated uncertainty \\
\hline
Deposition fraction ($D_F$) & $\pm$ 5\%  \\
Aspect Ratio ($\overline{L}$) & $\pm$ 1\%  \\ 
Dimensionless diameter ($\overline{D}$)  & $\pm$ 1.5\% \\
Reynolds number ($Re$) & $\pm$ 0.2\%  \\
Dimensionless Time ($\overline{T}$) & $\pm$ 0.6\%  \\
$\delta$ & $\pm$ 5\% \\ [0.5ex]
\hline \hline
\end{tabular}
\end{table}

\section{Effect of $\overline{L}$, $\overline{D}$ and $Re$ on deposition}

\begin{figure}[ht!]
\centering
\subfigure[]{\includegraphics[width=12cm,height=9.8cm]{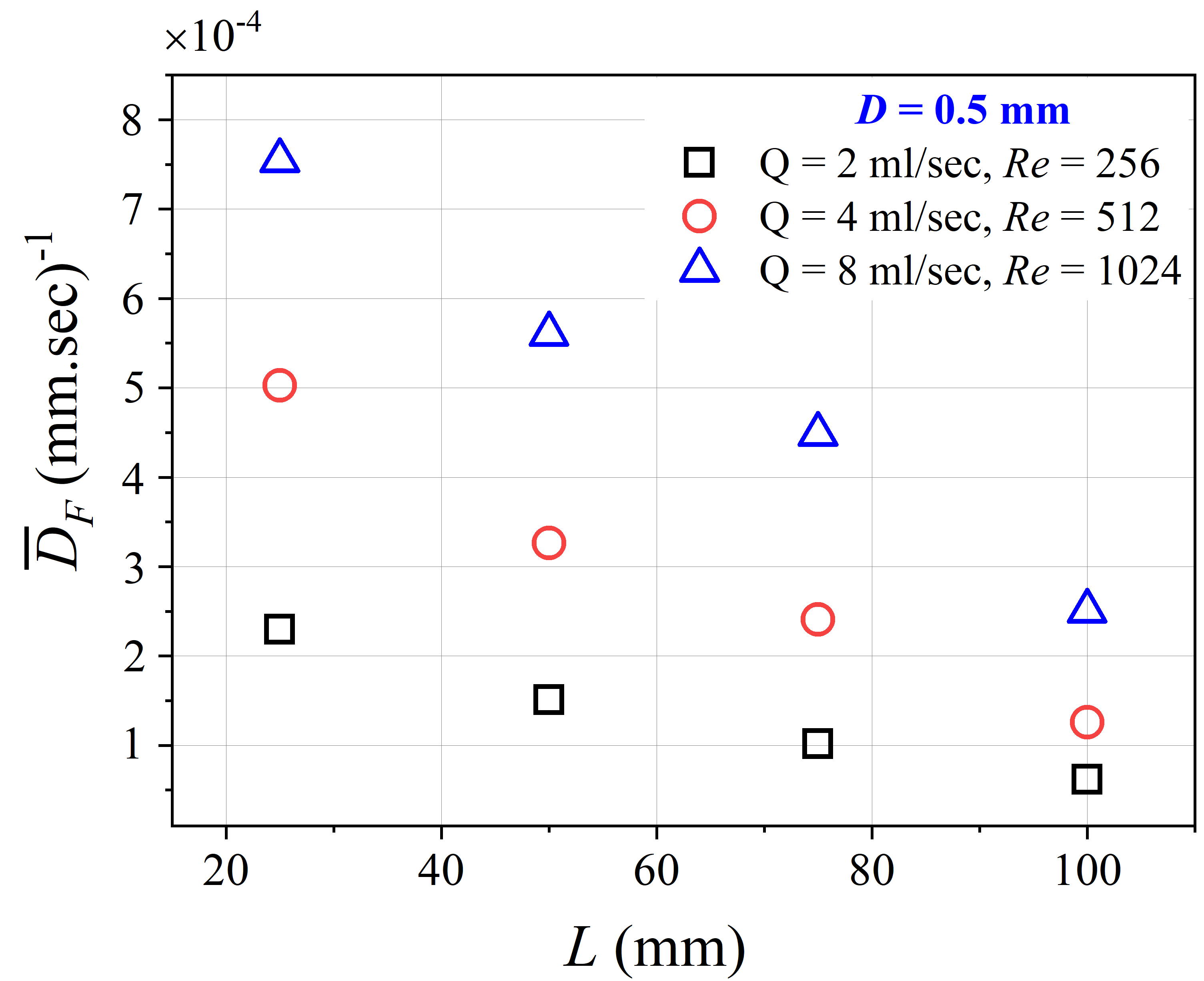}}
\subfigure[]{\includegraphics[width=12cm,height=9.5cm]{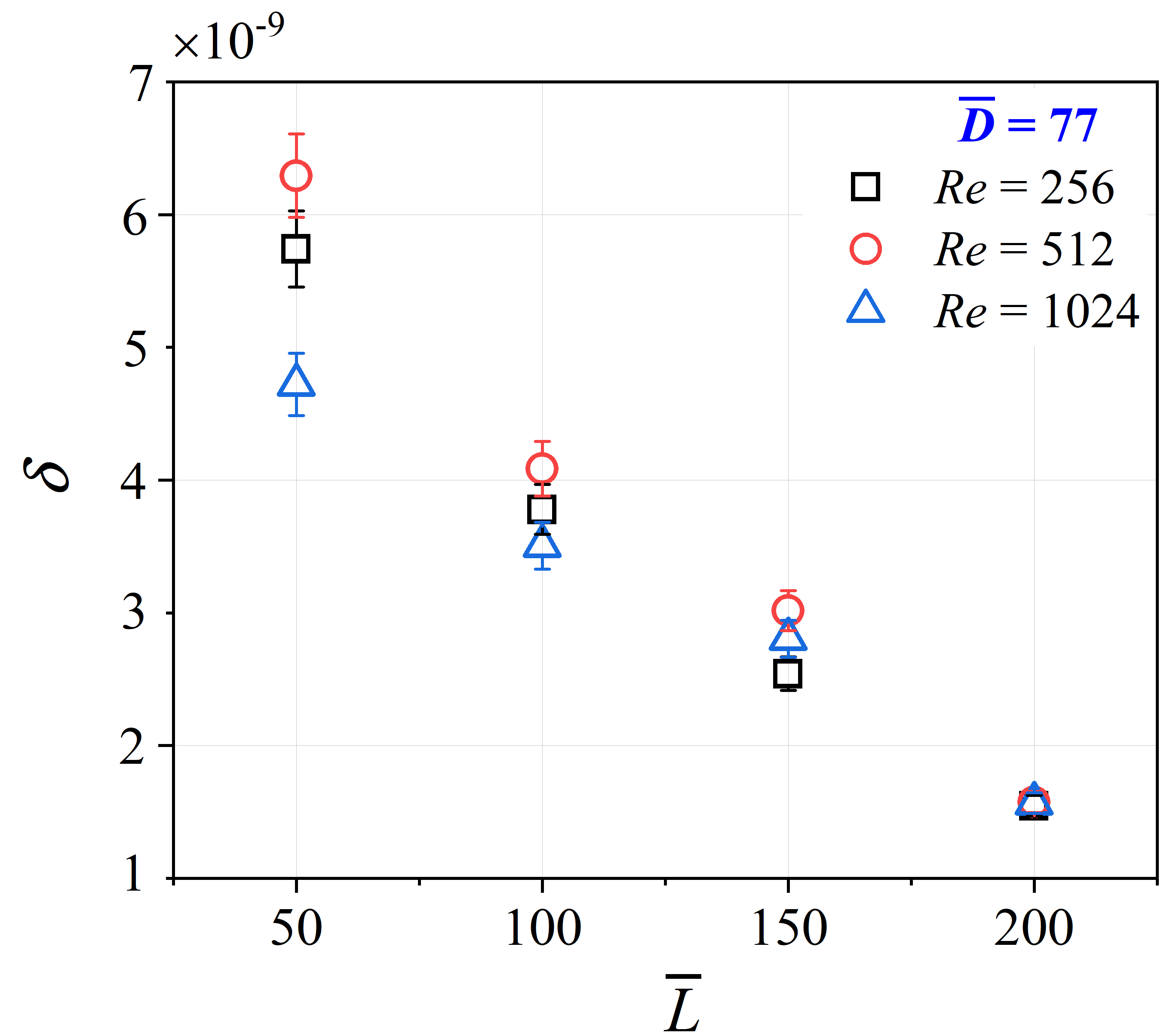}}
\caption{Plot of (a) variation of deposition fraction per unit capillary length per unit time ($\overline{D}_F$) with capillary length ($L$) and (b) the variation of dimensionless deposition ($\delta$) versus aspect ration ($\overline{L}$), for different Reynolds number ($Re$). The data shows an inverse relation of $\delta$ with bronchiole aspect ratio ($\overline{L}$). For high $\overline{L}$, the effect of $Re$ is negligible.}
\label{Figure: Constant D}
\end{figure}

The length of the capillary directly influences the aerosol deposition since it determines the aerosol travel time within the capillary. Figure \ref{Figure: Constant D}a shows that the $D_F$ per unit length per unit time is higher for small capillaries and it decreases as the capillary grow longer. Figure \ref{Figure: Constant D} (b) presents the variation of $\delta$ with $\overline{L}$ for different values of $Re$. The dimensional form of this variation is shown in figure \ref{Figure: Constant D} (a), where $\delta$ is replaced by deposition fraction ($D_F$) per unit length of the capillary, per unit time represented by $\overline{D}_F$. Within the lung, every bronchus is dichotomously branched into two bronchi at each generation. Thus the flow rate gets halved, which causes the $Re$ to become half of the previous value at every generation.The $Re$ is chosen such that it spans the range of Reynolds numbers encountered in the upper bronchi, where inertial effects are significant. $\overline{D}$ was maintained constant at $77$, which denotes the capillary diameter to be $0.5 mm$. The results show that $\delta$ decreases with an increase in $\overline{L}$. In other words, the deposition per unit length and unit time decreases as the length of the bronchiole increases (Ref: Figure \ref{Figure: Constant D}(a)). This shows the inverse relation of $\delta$ with $\overline{L}$ for all $Re$. For small aspect ratio, $\overline{L}$, the effect of $Re$ is visible. This is due to the fact that for small $\overline{L}$, inertial effects play a role in decreasing deposition. However, for high aspect ratios, the effect of $Re$ is insignificant. 

\begin{figure}[hbt!]
\centering
\subfigure[]{\includegraphics[width=12cm,height=9.5cm]{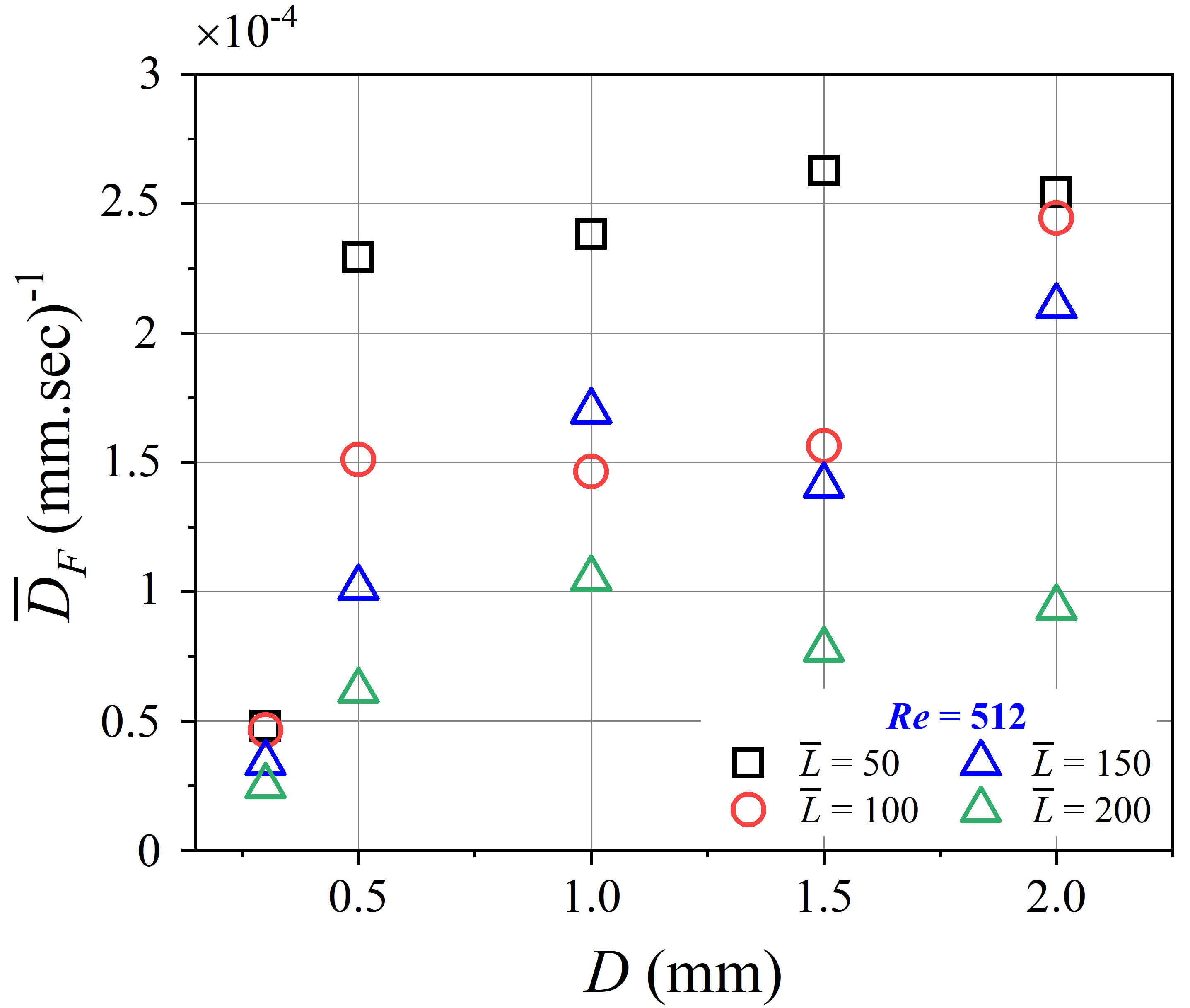}}
\subfigure[]{\includegraphics[width=12cm,height=9cm]{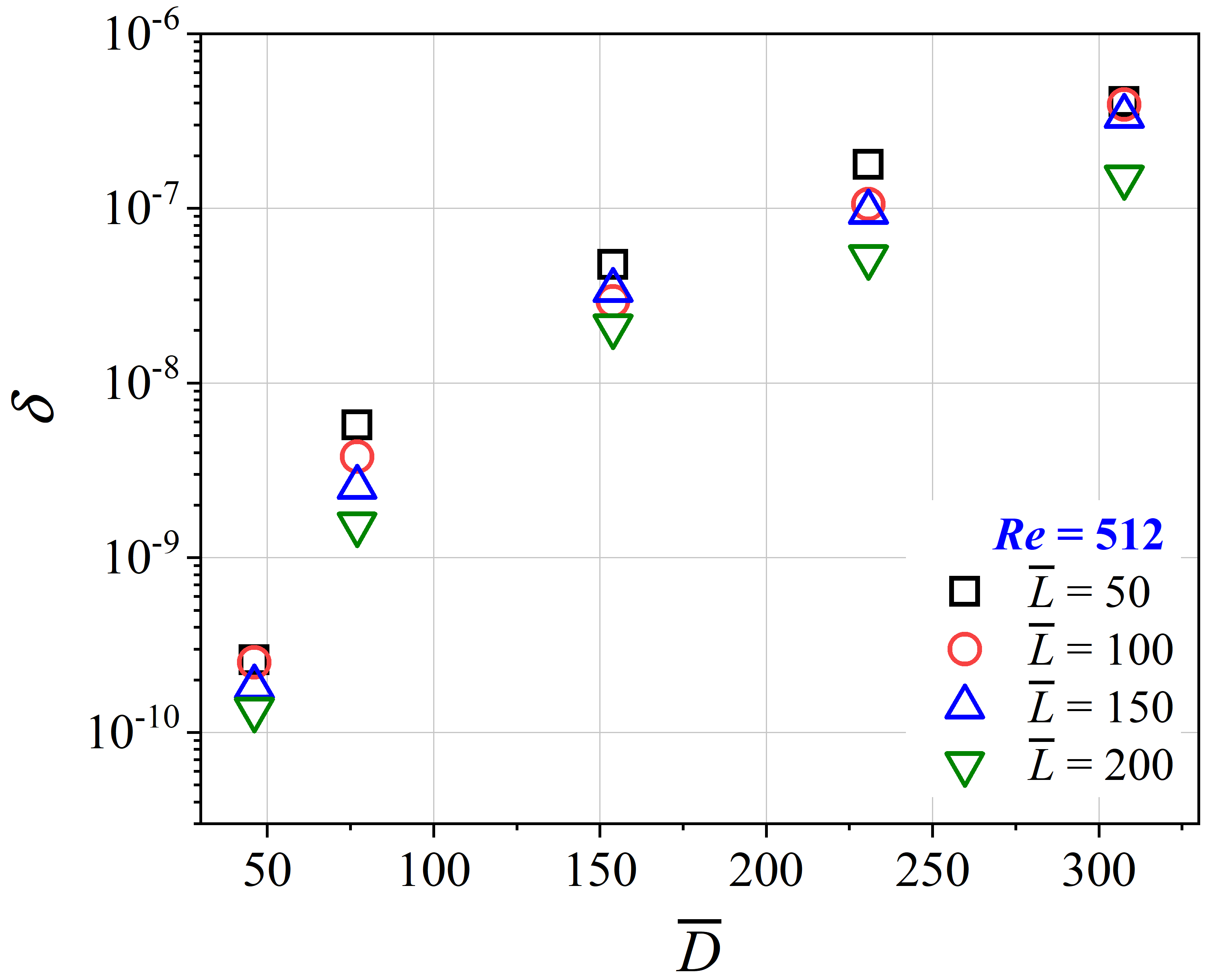}}
\caption{Plot of (a) deposition fraction per unit capillary length and unit time ($\overline{D}_F$) with tubules diameter ($D$) and (b) dimensionless deposition ($\delta$) versus dimensionless bronchiole diameter ($\overline{D}$), for varying bronchiole aspect ratio ($\overline{L}$). Reynolds number  was constant at $Re$ = 512. The tubules diameter mimics $7^{th} - 23^{rd}$ generation of the lungs. It is seen that with increase in $\overline{D}$ the dimensionless deposition ($\delta$) increases, while $\overline{L}$ still follows the inverse relation with $\delta$ for a constant $\overline{D}$.}
\label{Figure: Constant Re}
\end{figure}

\begin{figure}[hbt!]
\centering
\subfigure[]{\includegraphics[width=12cm,height=9.5cm]{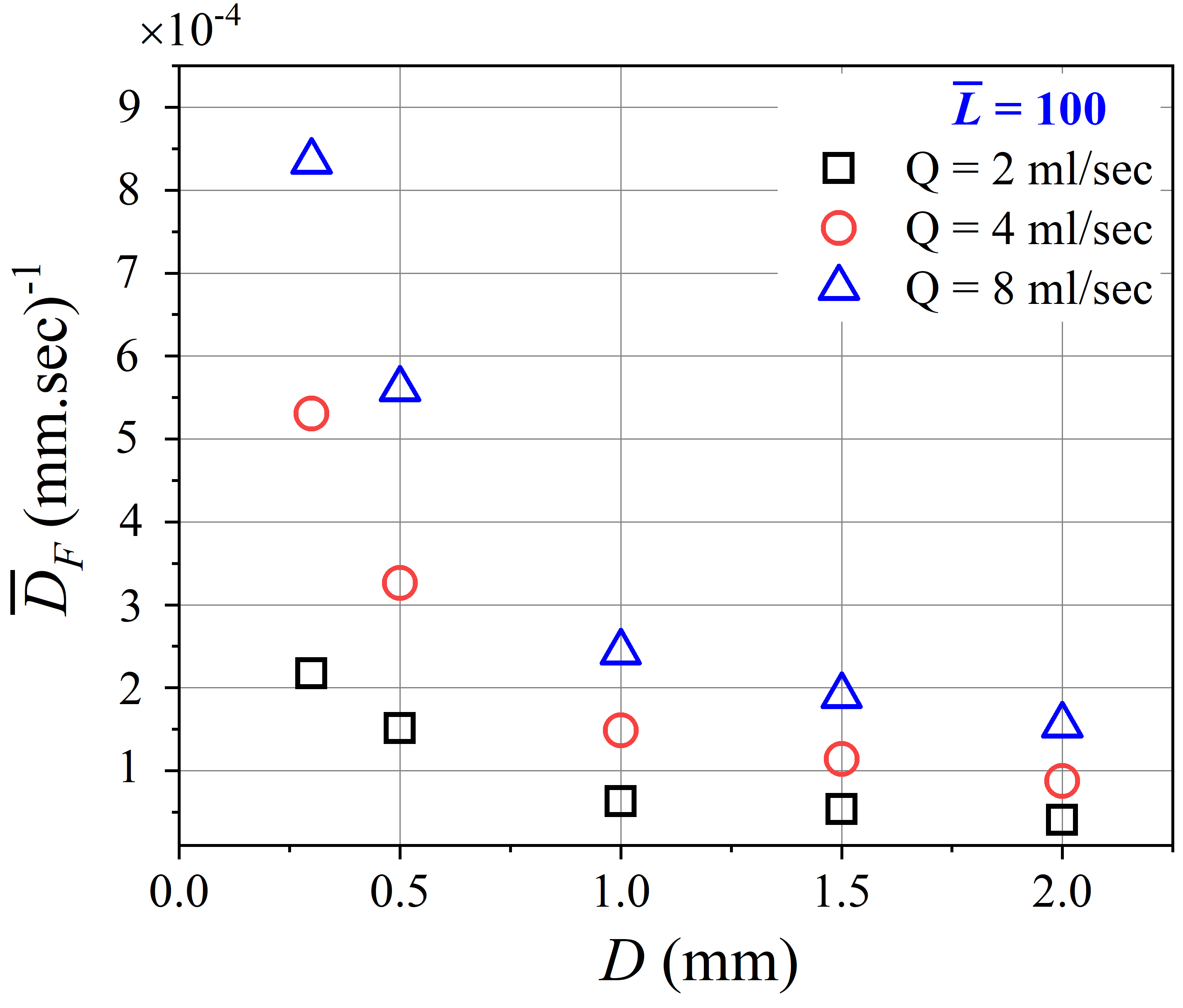}}
\subfigure[]{\includegraphics[width=12cm,height=9cm]{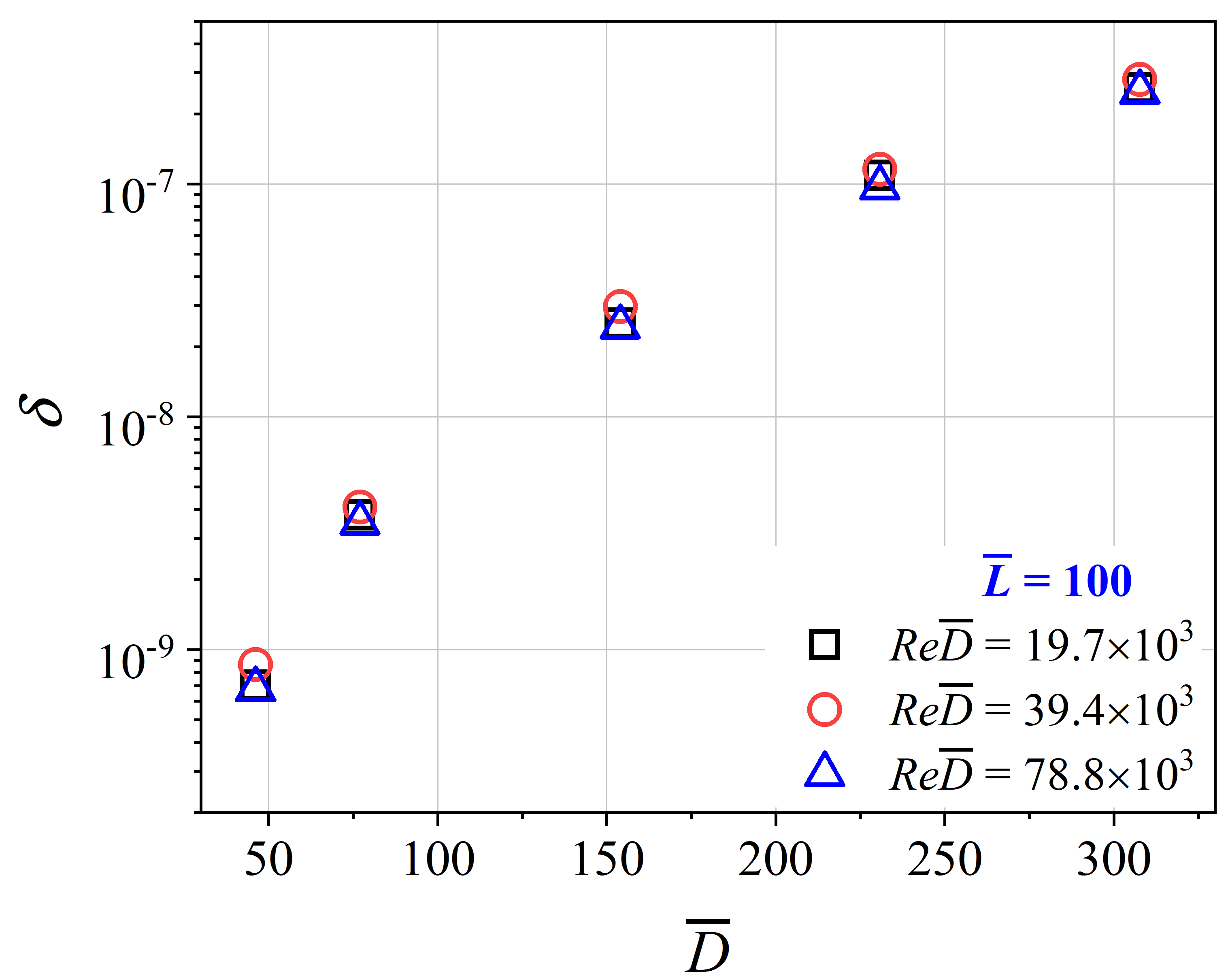}}

\caption{Plot of (a) deposition fraction per unit capillary length and unit time ($\overline{D}_F$) with tubules diameter ($D$) and  (b) dimensionless deposition ($\delta$) versus $\overline{D}$ for varying particle-based Reynolds number ($Re \overline{D}$), keeping aspect ratio $\overline{L} = 100$. It is seen that $\delta$ increases with $\overline{D}$. $Re \overline{D}$ does not affect the deposition.}
\label{Figure: Constant L}
\end{figure}

The effect of particle size on regional deposition has been studied in the past literature \citep{lippmann1969effect,mitchell1987effect,glover2008effect,malve2020modelling} but the effect of bronchiole diameter on aerosol deposition is least studied. With different generations of the lungs the bronchiole diameter varies continuously and affects deposition.
The effect of ${\overline{D}_F}$ for different capillary diameter ranging from $0.3 mm - 2 mm$ is shown in figure \ref{Figure: Constant Re} (a), for constant $Re = 512$. The diameter of the tubules mimics $7^{th}$ to $23^{rd}$ generations of the lungs. The effect of dimensionless bronchiole diameter on deposition ($\delta$) is shown in Figure \ref{Figure: Constant Re} (b) for various aspect ratios ($\overline{L}$), maintaining $Re$ constant. $\delta$, representing deposition fraction per unit length of the bronchiole per unit suction time, increases with an increase in bronchiole diameter, represented by $\overline{D}$. The rate of increase of $\delta$ decreases for $\overline{D}>150$. This is because for a constant $Re$, an increase in bronchiole diameter causes the flow rate to increase which results in higher deposition. The value of $\delta$ is lowest for $\overline{L} = 200$ and increases with decrease in $\overline{L}$ similar to the trends in figure \ref{Figure: Constant L} (b).

The particle deposition in lungs greatly depends on the airflow rate \citep{zhang2000effects,koullapis2018effect}. Although airflow rate can be represented non dimensionally by flow $Re$ but $Re$ changes with diameter even if the flow rate is constant. In a real case scenario, each generations of lung may undergo different air flow rates based on different breathing conditions \citep{deng2019particle}. 
Figure \ref{Figure: Constant L} (a) shows the plot of $\overline{D}_F$ for variable tubule diameter and air flow rates. The aspect ratio is kept constant, $\overline{L} = 100$. At each lung generation the airways gets bifurcated causing the flow rate to be halved. Thus in the experiment the flow rates increases twice the previous value. Figure \ref{Figure: Constant L} (b) is the plot of $\delta$ versus $\overline{D}$ for different flow conditions. The suction flow rates are represented non-dimensionally by a particle-based Reynolds number $Re_p=4Q/(\pi \nu D_{10})$. Interestingly, $Re_p=Re\overline{D}$, which can be understood as a Reynolds number based on the mean particle size, $D_{10}$. As can be seen, $\delta$ increases with an increase in the dimensionless bronchiole diameter ($\overline{D}$), but is not dependent on the particle based Reynolds number. This is unlike the data presented for a constant $Re$ in Figure \ref{Figure: Constant Re} (b). As $\overline{D}$ increases at a constant $Re \overline{D}$, the mean velocity of the air flow in the bronchiole decreases. This causes the rate of deposition to increase, since diffusion-driven and gravitational settling become relevant. It is important to note that both Figures \ref{Figure: Constant Re} (b) and \ref{Figure: Constant L} (b) are plotted with the $\delta$ co-ordinate being plotted on a logarithmic axes. A factor of $6$ change in $\overline{D}$ brings about three orders of magnitude change to the deposition fraction. Therefore, it can be concluded that the bronchiole diameter is an important parameter in determining aerosol deposition. 

\clearpage
\section{Deposition for the entire lung-specific $Re$ range}

The tidal volume of a healthy adult human lung is approximately $500 ml$ or $7 kg$ / body mass per inspiration \citep{ricard2003we,mackay2020computed}. The tidal volume is described as the volume of air delivered to the lung with each breath when no extra effort is applied. Considering the flow of the air to be $500 ml$ for a breathing cycle of 4 seconds (i.e. 2 seconds each for inspiration and expiration) \citep{fox2015human}, the Reynolds number ($Re$) in trachea ($G = 0$) and alveoli ($G = 23$) are respectively, on the $\mathcal{O}(10^3)$ and $\mathcal{O}(10^{-2})$ (Ref: Figure \ref{Figure: Re of different gen}). This huge variation of $Re$ within different generations of the lung is responsible for different modes of aerosol deposition in different regions, such as, deposition by impaction, sedimentation and diffusion \citep{grotberg2011respiratory,rostami2009computational,cheng2014mechanisms,svartengren1987regional}.
\par
\begin{figure}[hbt!]
\centering
\subfigure[]{\includegraphics[width=11cm,height=8cm]{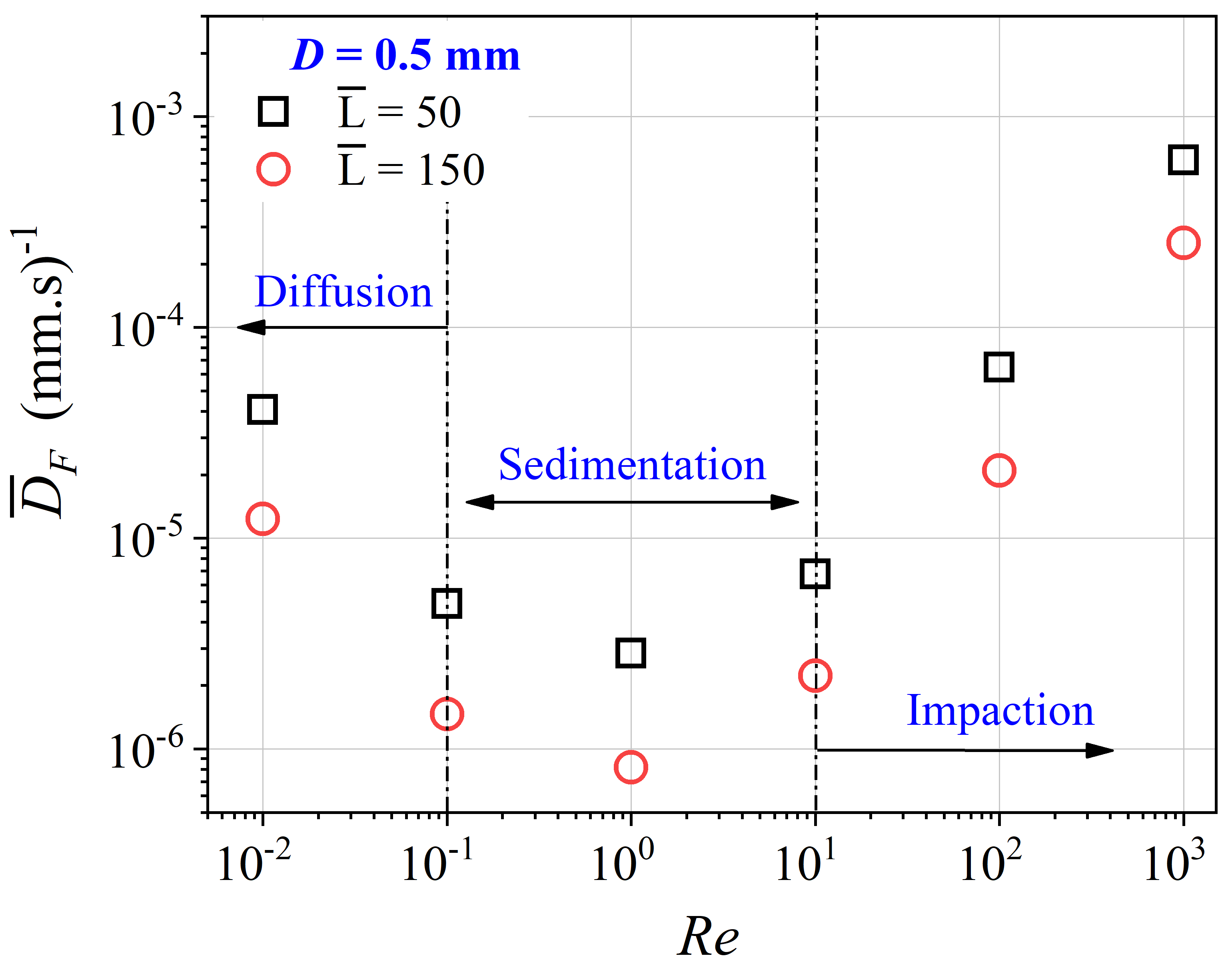}}
\subfigure[]{\includegraphics[width=11cm,height=8cm]{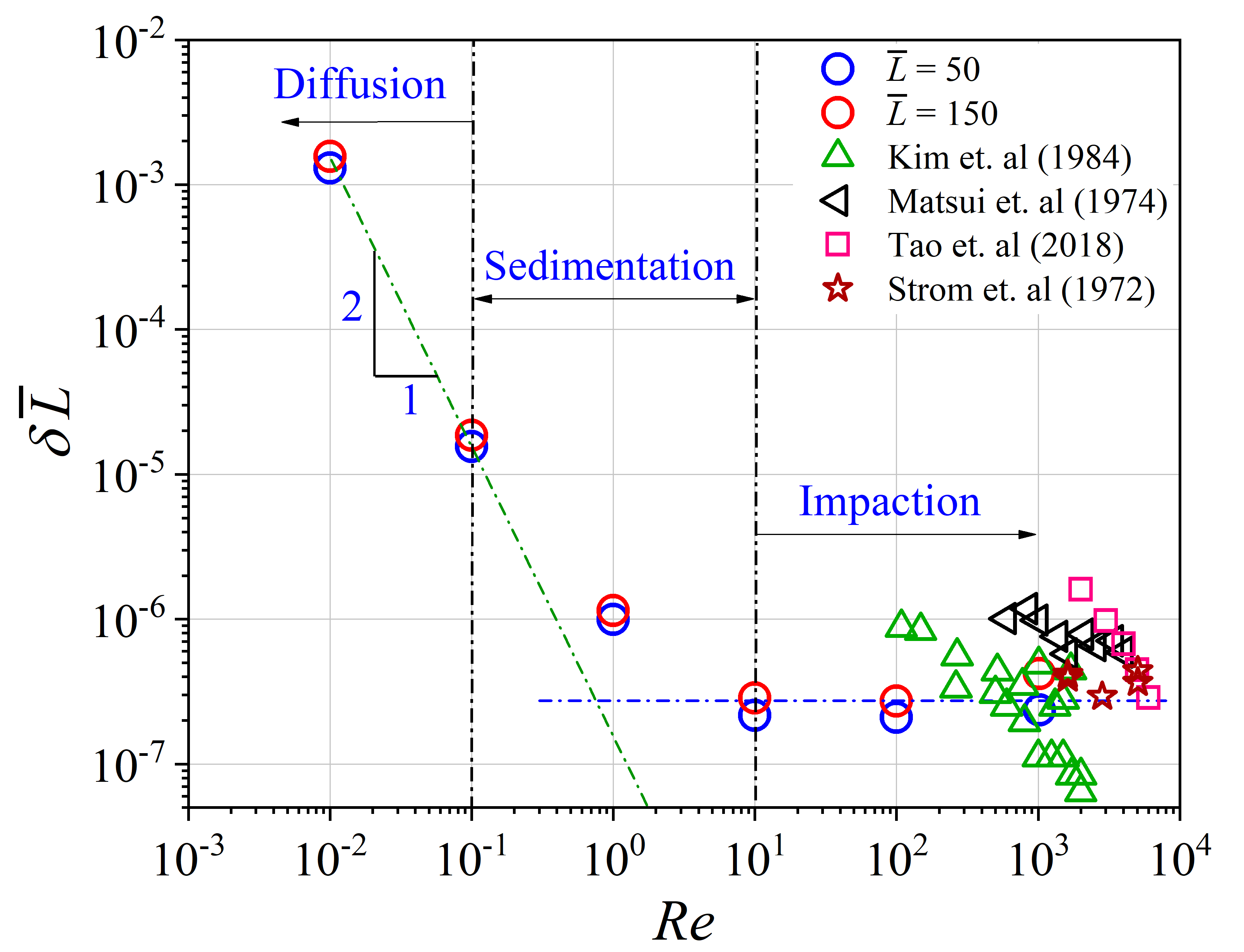}}
\caption{(a) Plot of $\overline{D}_F$ versus $Re$. $Re$ varies over five orders of magnitude of mimicking flow conditions in different generations of the lung. The deposition fraction per unit tubule length and per unit suction time represented by $D_F$ shows the lowest value when $Re = 1$, for $\overline{L} = 50$ and $150$. According to the deposition mechanism of aerosols three zones are identified as follows: deposition by \textit{impaction} ($10 < Re \leqslant 10^3$), deposition by \textit{sedimentation} ($10 < Re < 1$), deposition by \textit{diffusion} ($10^{-3} \leqslant Re < 1$). (b) Plot of $\delta \overline{L}$ versus $Re$ for $\overline{D}$ being constant at $77$ and for $\overline{L} = 50$ and $150$. $Re$ varies over five orders of magnitude. The data in figure \ref{Figure: Low Re} (a) collapses onto a single curve when re-scaled as above. Interestingly, $\delta \overline{L} \sim Re^{-2}$ for $Re < 1$ and is constant for $Re > 1$. In the plot green dotted line is the best fit for $Re < 1$ given by $\delta \overline{L} = 1.55 \times 10^{-7}Re^{-2}$. Blue dotted line is the best fit for $Re>1$ given by $\delta \overline{L} = 2.73 \times 10^{-7}$. Figure legend:  \textcolor{Green}{$\triangle$} \citet{kim1984deposition}; \textcolor{black}{$\lhd$} \citet{Matsui1974}; \textcolor{Magenta}{$\square$} \citet{Tao2018}; \textcolor{brown}{\ding{73}} \citet{strom1972transmission}}
\label{Figure: Low Re}
\end{figure}

In this study the experiments were carried out for different orders of $Re$ ranging from $10^{-2} \leqslant Re \leqslant 10^3$, representing the entire lungs. The existing literature speaks about the total deposition in lungs,numerically \citep{kolanjiyil2016computationally,kolanjiyil2017computationally,kolanjiyil2017computational, brand2000total,jaques2000measurement,madureira2020assessment,dolovich2001measuring}. To the best of our knowledge, this is the first report where the entire dynamic range of lung-relevant Reynolds numbers have been explored in one experiment.  Figure \ref{Figure: Low Re} (a) represents the plot of $\overline{D}_F$ for entire range of $Re$, that happens in human lungs in a single breath. A minimum in $D_F$ can be observed nearly at $Re = 1$, where the velocity is in $\sim \mathcal{O}(10^{-1})$ $m/s$ and gravitational deposition dominates over inertial deposition. For impaction dominated zone (i.e. $10 \leqslant Re \leqslant 10^3$) and in diffusion dominated zone ($10^{-2} \leqslant Re \leqslant 1$), $D_F$ takes on higher values. This is because a high suction rate carries greater number of aerosol particles that impact and deposit on the bronchiole walls. In the diffusion-dominated zone again, $D_F$ increases with a decrease in $Re$. This is possibly due to the fact as $Re$ decreases, thermal fluctuations become dominant in moving the aerosol towards bronchiole walls. Thus for a sufficiently long exposure time, the aerosol diffuses inside the bronchiole causing increased deposition.
\par
The effect of $Re$ on $\delta \overline{L}$ is investigated for $\overline{L} = 50$ and $150$ in Figure \ref{Figure: Low Re} (b). The experiment has been carried out for two aspect ratios, to verify the inverse relation between $\delta$ and $\overline{L}$ still follows at low $Re$ condition. It may be recalled that $\delta \overline{L}$ is the dimensionless rate of deposition. It can be seen that when $\delta \overline{L}$ is plotted against $Re$, the data in Figure \ref{Figure: Low Re} (b) nicely collapse onto a single curve. This data collapse points to a minimal set of dimensionless parameters that is required to completely describe aerosol deposition. Two asymptotic regimes can be seen in Figure \ref{Figure: Low Re} (b). For $Re > 1$, $\delta \overline{L}$ is independent of $Re$. This is the parametric regime where deposition happens mostly due to impaction on the bronchiole walls. For $Re<1$, $\delta \overline{L} \sim Re^{-2}$. For $Re = 1$ the velocity of the flow is in $\sim \mathcal{O}(10^{-2})$ $m/s$ which causes sedimentation of particles and increases $\delta$ by an order of magnitude. Further reduction of $Re$ causes the suction velocity to be small enough that diffusion becomes the dominant mode of deposition. In addition, $\delta \overline{L}$ increases several orders of magnitude as $Re$ is decreased from $10^3$ to $10^{-2}$ i.e. from the impaction regime to the diffusion regime. The order of $\delta \overline{L}$ is almost constant in the impaction regime while it increases for $Re < 1$. Data from the literature has also been re-plotted in the current nomenclature in this figure. Firstly, it can be seen that at any value of $Re$, there is significant variation in the data from the literature. Secondly, the data in the literature is limited to values of $Re>10^2$. It can be seen from the figure \ref{Figure: Low Re} (b), that our data is within the range of values from the literature. This can be noted as the most important finding of this study that is hardly available in the existing literature.

\chapter{Deposition Modelling}

This chapter discusses the implications of the functional form of the deposition correlation $G$, from the experimental results for the equation \ref{eqn:dimrel}. This is a physics-consistent, dimensionless deposition model that can be used to estimate the regional deposition in lungs for a given breathing condition and particle size.

\section{Model development}
\begin{figure}[hbt!]
\centering
\subfigure[]{\includegraphics[width=12cm,height=9cm]{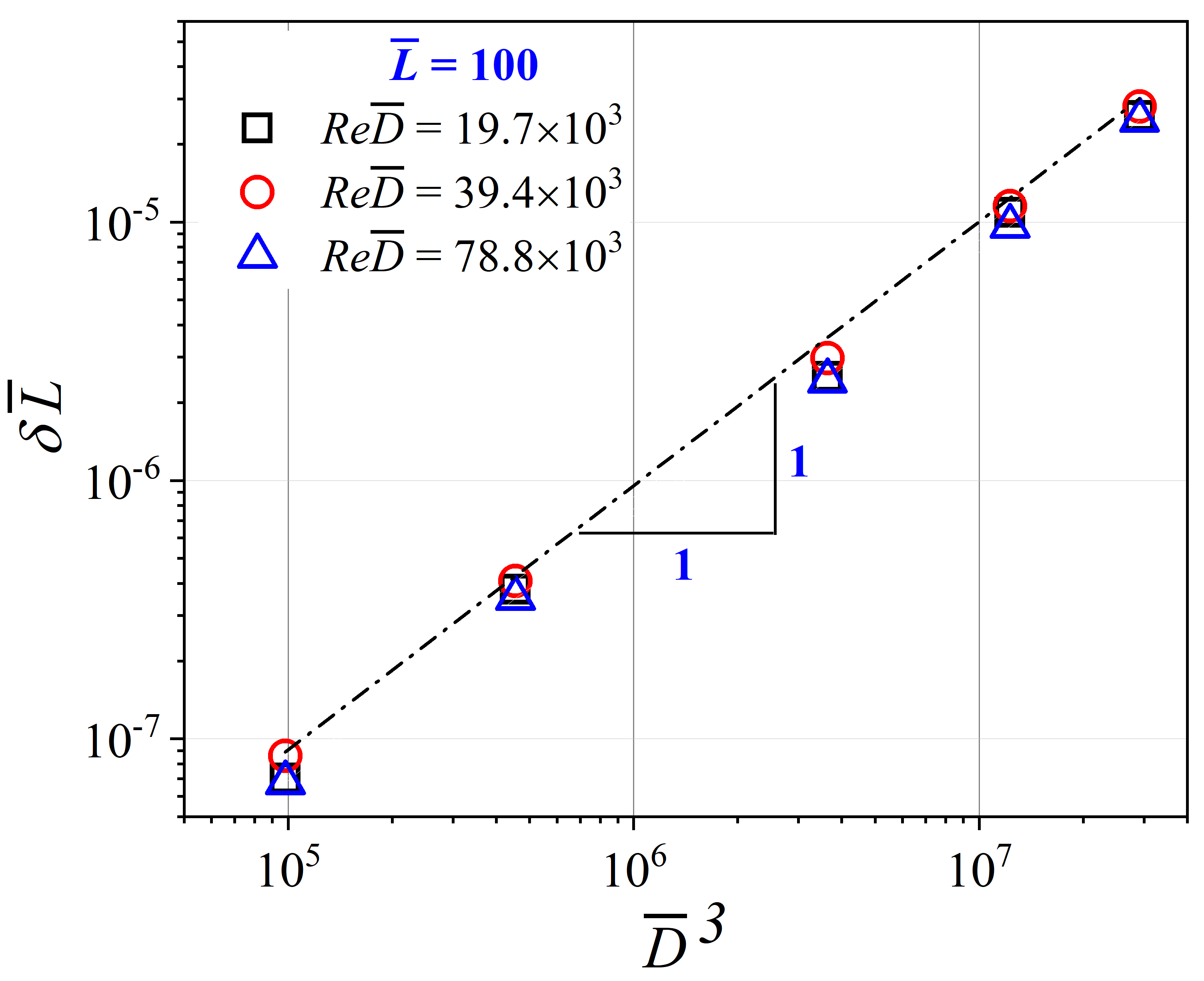}}
\subfigure[]{\includegraphics[width=12cm,height=9cm]{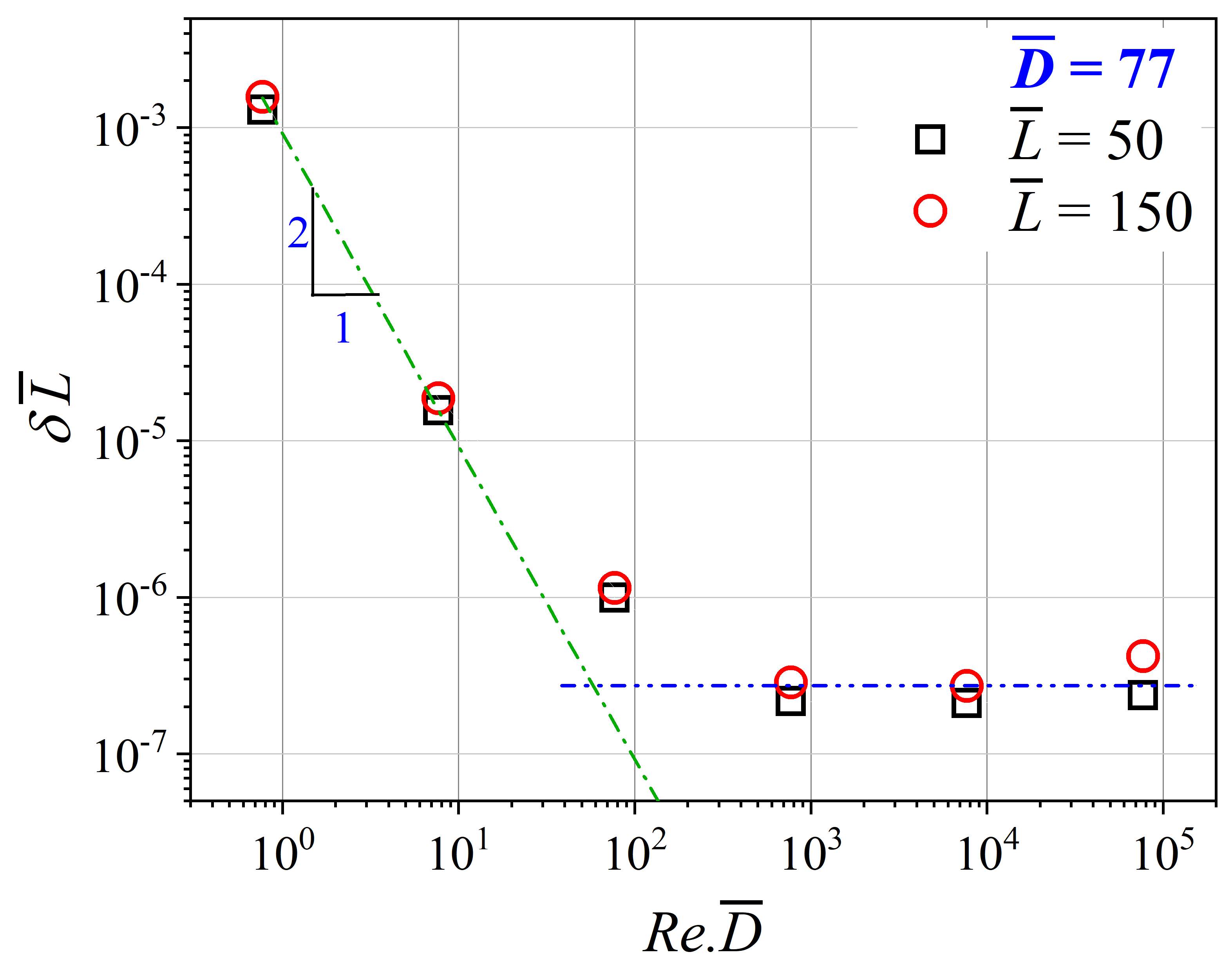}}
\caption{(a) Modified representation of figure \ref{Figure: Constant L}(b). The same data is re-plotted to show the variation of $\delta \overline{L}$ versus $\overline{D}^{3}$. The best fit power law is given by $\delta \overline{L} = 7 \times 10^{-13}\overline{D}^{3}$.  The fit is independent of $Re\overline{D}$. $Re\sim 10^3$ for all data in these plots implying impaction-dominated deposition. (b) Modified representation of figure \ref{Figure: Low Re}(b) where $\delta\overline{L}$ is plotted against particle Reynolds number ($Re\overline{D}$ instead of flow $Re$. The best fit represented by green dotted line is given by $9 \times 10^{-4} (ReD)^{-2}$ }
\label{Figure: Ranges of Re for dL}
\end{figure}

A dimensionless deposition model is developed from the experimental results which is the main focus of this study. All the parameters influencing the deposition is considered in this study and non dimensional form of those parameters are used to develop the model so that the model can be used for all dimensional forms.
Figures \ref{Figure: Ranges of Re for dL}(a) and \ref{Figure: Ranges of Re for dL}(b) are modified representations of figures \ref{Figure: Constant L}(b) and \ref{Figure: Low Re}(b) respectively. These figures represent the same phenomena as from the earlier figures but in terms of modified dimensionless parameters. Figure \ref{Figure: Ranges of Re for dL}(a) shows that $\delta \overline{L}$ linearly scales to $\overline{D}^3$ and it is independent of particle Reynolds number ($Re \overline{D}$). In this case, $Re \gg 1$. Similar result is obtained from figure \ref{Figure: Low Re}(b) as well as figure \ref{Figure: Ranges of Re for dL}(b) where $\delta \overline{L}$ is independent of $Re \overline{D}$ for $Re \gg 1$. But for $Re \ll 1$, $\delta \overline{L}$ strongly depends on particle Reynolds number ($Re \overline{D}$) as indicated in figure \ref{Figure: Ranges of Re for dL}(b). The collapse of data in Figures \ref{Figure: Ranges of Re for dL}(a) and \ref{Figure: Ranges of Re for dL}(b) points to a universal description of deposition in dimensionless terms as a function of $\overline{D}$ and $Re$, the two abscissa parameters in these figures. The best fit to the data in figure \ref{Figure: Ranges of Re for dL}(a) suggests that $\delta \overline{L} \sim \overline{D}^3$ for $Re \gg 1$ and independent of $Re$. Therefore, one can related $\delta\overline{L}$ to $\overline{D}$. A curve fit for the data pertaining to $Re \ll 1$ in figure \ref{Figure: Ranges of Re for dL}(b) suggests that $\delta \overline{L} \sim (Re\overline{D})^{-2}$. By combining the two relations obtained from figures \ref{Figure: Ranges of Re for dL}(a) and (b), a universal deposition equation is derived as follows.
\begin{subequations}\label{eq: Best Fit:main}
\begin{align}
\delta \overline{L}&=7 \times 10^{-13} \overline{D}^3 &\text{if} \hspace{4pt} Re \gg 1 \label{eq: Best Fit:a}\\
&=1.55 \times 10^{-7} Re^{-2} &\text{if} \hspace{4pt} Re \ll 1 \label{eq: Best Fit:b}
\end{align}
\end{subequations}
\noindent The equations \ref{eq: Best Fit:main} explain the data in both the figures \ref{Figure: Ranges of Re for dL}(a) and (b). This is the explicit form of equation \ref{eqn:dimrel} that we set out to identify. Equations \ref{eq: Best Fit:main} also indicate the minimal set of dimensionless parameters that are required to model aerosol deposition over the entire range of lung-relevant operating conditions.

In order to understand the physics underlying equations \ref{eq: Best Fit:main}, it is useful to recast equations \ref{eq: Best Fit:main} back into the dimensional parameter form. From the dimensional form of equation \ref{eq: Best Fit:a} for $Re \ll 1$, we find that $d \propto {\nu^2 T^2 \alpha_0}/{D_{10}}$. Recall that $d$ is defined as the total volume of aerosol deposited in time $T$. For $Re \ll 1$, as expected, the flow rate and other parameters do not play a role. The deposition depends linearly on the particle concentration $\alpha_0$. The rate of deposition is proportional to $D_{10}^{-1}$ as one would expect in diffusion-dominated deposition (since the diffusion co-efficient scales as $D_{10}^{-1}$). From the dimensional form of equation \ref{eq: Best Fit:b} for $Re \gg 1$, we find that $d \propto {Q^2T^2\alpha_0}/{D_{10}^3}$. Again, as expected, the deposition in this case is dependent on the square of the velocity ($Q^2$) and depends linearly on the particle concentration ($\alpha_0$), since $d$ is dominated by impaction. 
\par
\begin{figure}[hbt!]
\centering
\subfigure[]{\includegraphics[width=12cm,height=9cm]{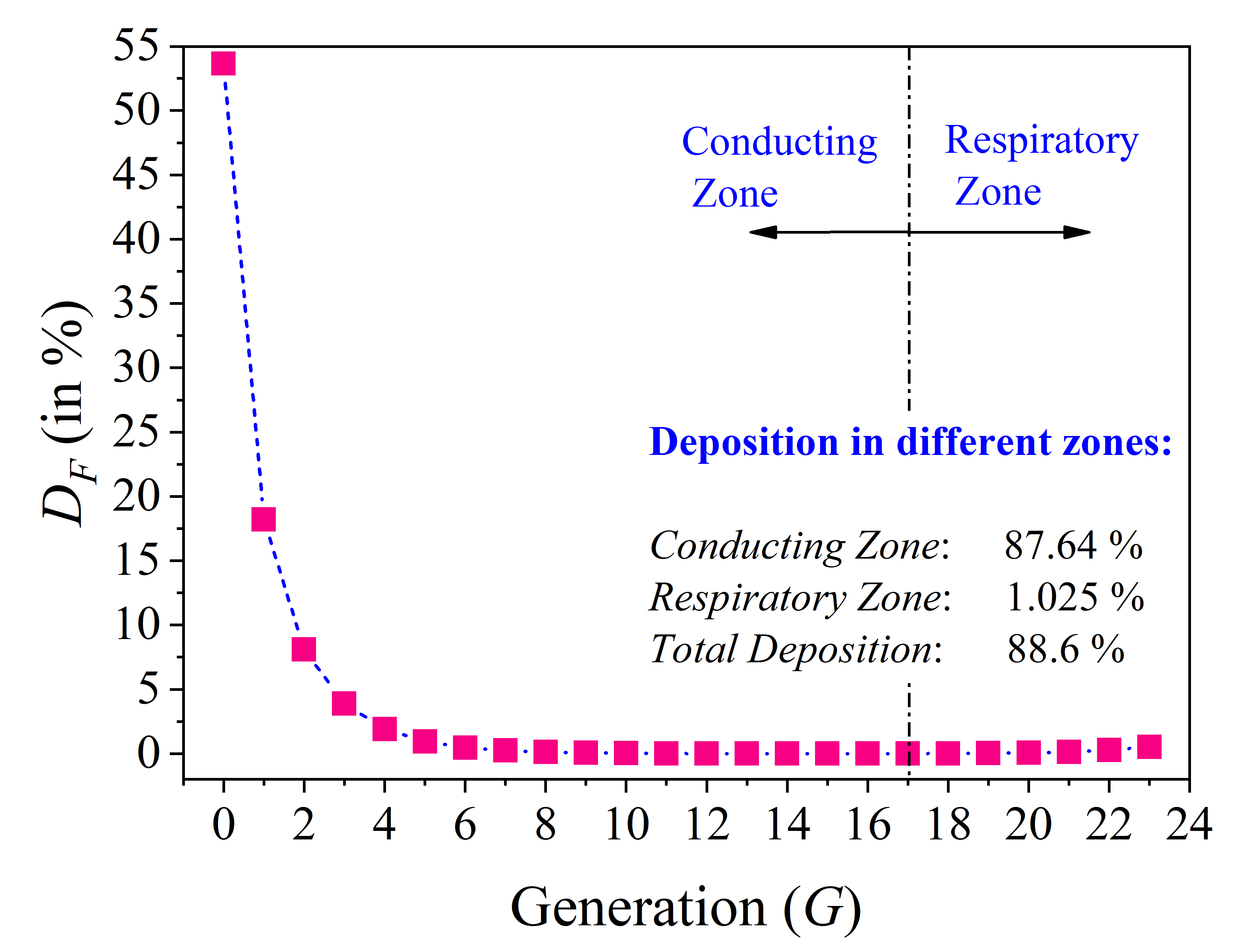}}
\subfigure[]{\includegraphics[width=12cm,height=9cm]{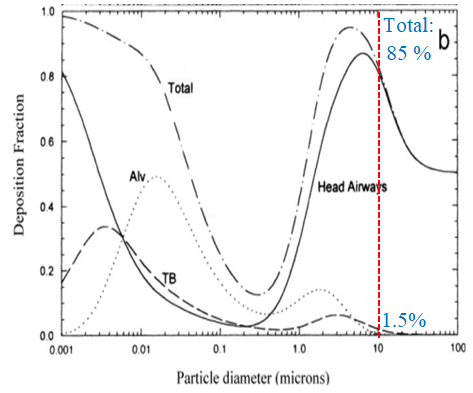}}
\caption{(a) Plot of deposition fraction ($D_F$) versus lung generations ($G$) calculated from equation \ref{eq: Best Fit:main} for particle size $\sim \mathcal{O}(10 \mu m)$ and breathing cycle of 4s. (b) Variation of deposition fraction with particle diameter given by \cite{hinds1999aerosol}. The plot shows that for particle diameter of 10 micron the total deposition fraction is 83\%, out of which 2\% is deposited in alveolar and tracheobroncial region and remaining in the upper airways. This results matches well with model predicted deposition in plot (a) }
\label{Figure: model validation}
\end{figure}

The experimentally developed model is validated for breathing cycle of 4 seconds and $D_{10} \sim \mathcal{O}(10 \mu m)$ with the results obtained by \cite{hinds1999aerosol}. The figure \ref{Figure: model validation}(a) shows the deposition for different lungs generations estimated from the model. It is calculated from the model that among total deposition of $88.6\%$, the deposition in conducting zone is $87.64\%$ and in respiratory zone is $1.025\%$ which matches well from the result of \cite{hinds1999aerosol} as shown in figure \ref{Figure: model validation}(b).

\clearpage

\section{Deposition calculation using the Model}
\begin{figure}[hbt!]
\centering
{\includegraphics[width=12cm,height=9cm]{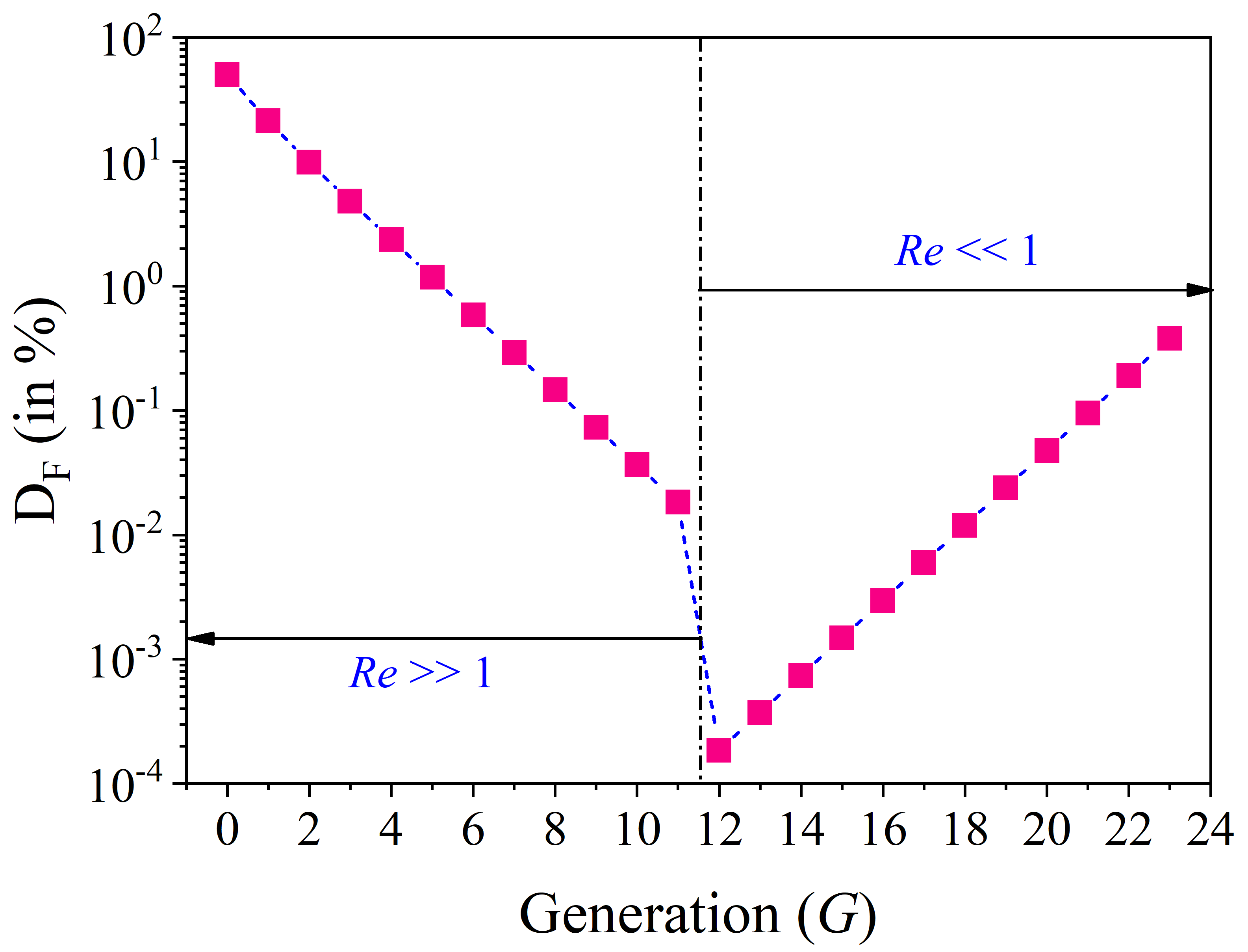}}

\caption{Plot of deposition fraction ($D_F$) and generation number, calculated using equation \ref{eq: Best Fit:main} for single breathing cycle (4s) and particle size $\sim \mathcal{O}(10 \mu m)$ The plot is same as figure \ref{Figure: model validation}(a) with the abscissa in the log scale. The discontinuity of the plot after $G = 11$ is due to the change in deposition model from equation \ref{eq: Best Fit:main}(a) to (b) resulted due to the change in $Re$. }
\label{Figure: Regional deposition from model}
\end{figure}

The intention of developing the model from the experimental results is to calculate the regional deposition in lungs for estimating the efficacy of the delivered drug. The deposition fraction is calculated using the model (ref: figure \ref{Figure: Regional deposition from model} for single breathing cycle of 4 seconds (inspiration: 2s; expiration: 2s) and particle size $\sim \mathcal{O}(10 \mu m)$, which is common among the commercially available nebulizers. 

In the figure \ref{Figure: Regional deposition from model} the deposition decreases exponentially from $0 \leqslant G \leqslant 11$ and rising exponentially for $12 \leqslant G \leqslant 23$. The discontinuity between generation 11 and 12 is due to the change of the deposition model. The deposition model in equation \eqref{eq: Best Fit:main} is valid in the asymptotic limits of $Re \ll 1$ and $Re \gg 1$. The region near $Re \sim 1$ is a cross-over region where the equations are strictly not valid. This is the source of the discontinuity in Figure \ref{Figure: Regional deposition from model}. The figure shows a total deposition fraction of 82.5\% out of which 81.9\% deposits in the air conducting zone and 0.6\% is deposited in the respiratory zone of the lungs. This low deposition fraction in the respiratory zone i.e. $17 \leqslant G \leqslant 23$ reduces the efficacy of the drug since most of the inhaled particles deposits in the upper lungs where it cannot be introduced into the blood stream. Thus the equation \ref{eq: Best Fit:main} is a closed form expression to estimate the deposition in any bronchiole in the lung, given its geometry and flow properties. We have ignored orientation of the bronchiole, since we have only studied deposition in horizontal bronchioles. However, as pointed out by \cite{GoldbergSmith}, one could account for the orientation angle by re-scaling time using the cosine of the angle of inclination.
\par
\begin{figure}[hbt!]
\centering
\subfigure[]{\includegraphics[width=12cm,height=9cm]{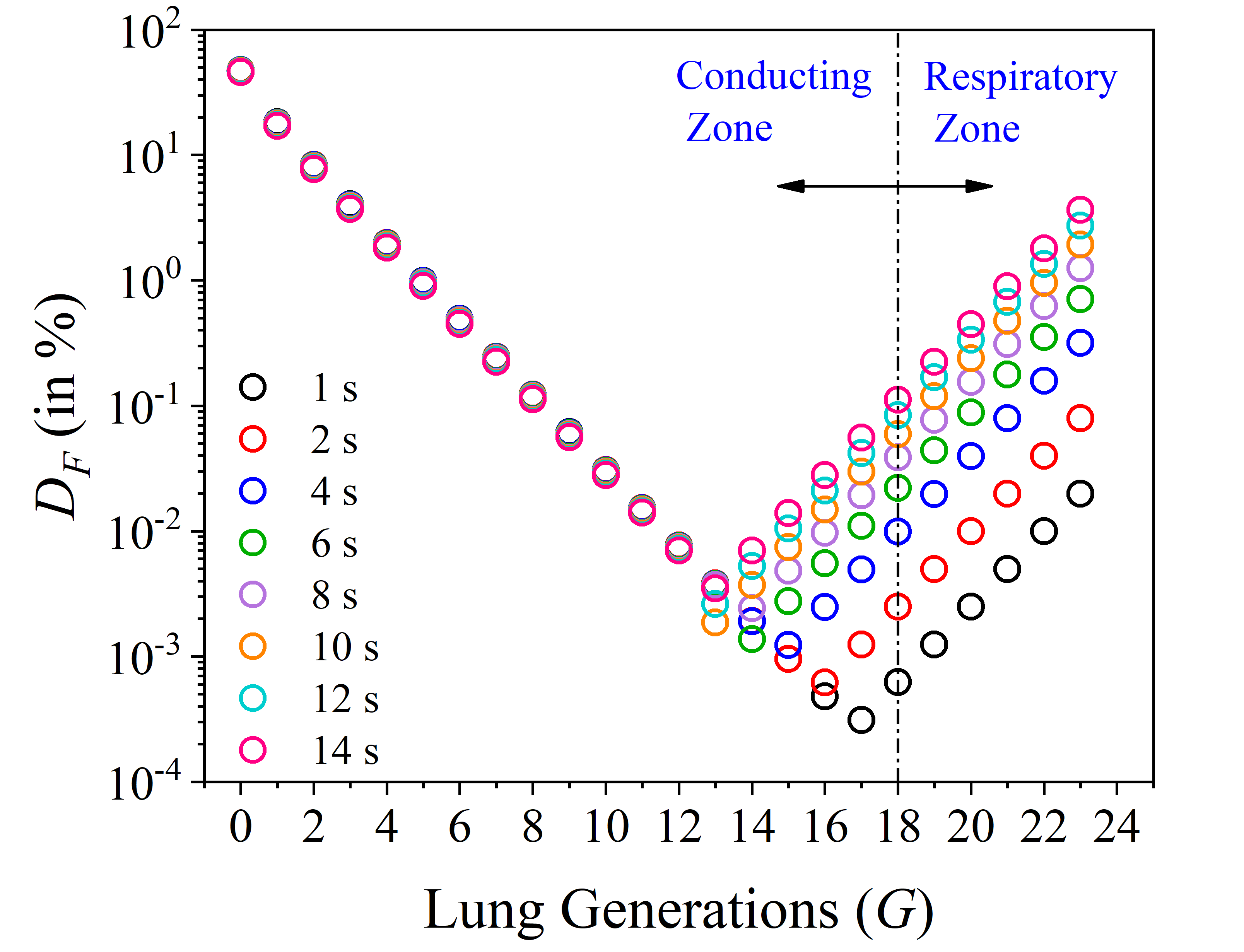}}
\subfigure[]{\includegraphics[width=12cm,height=9cm]{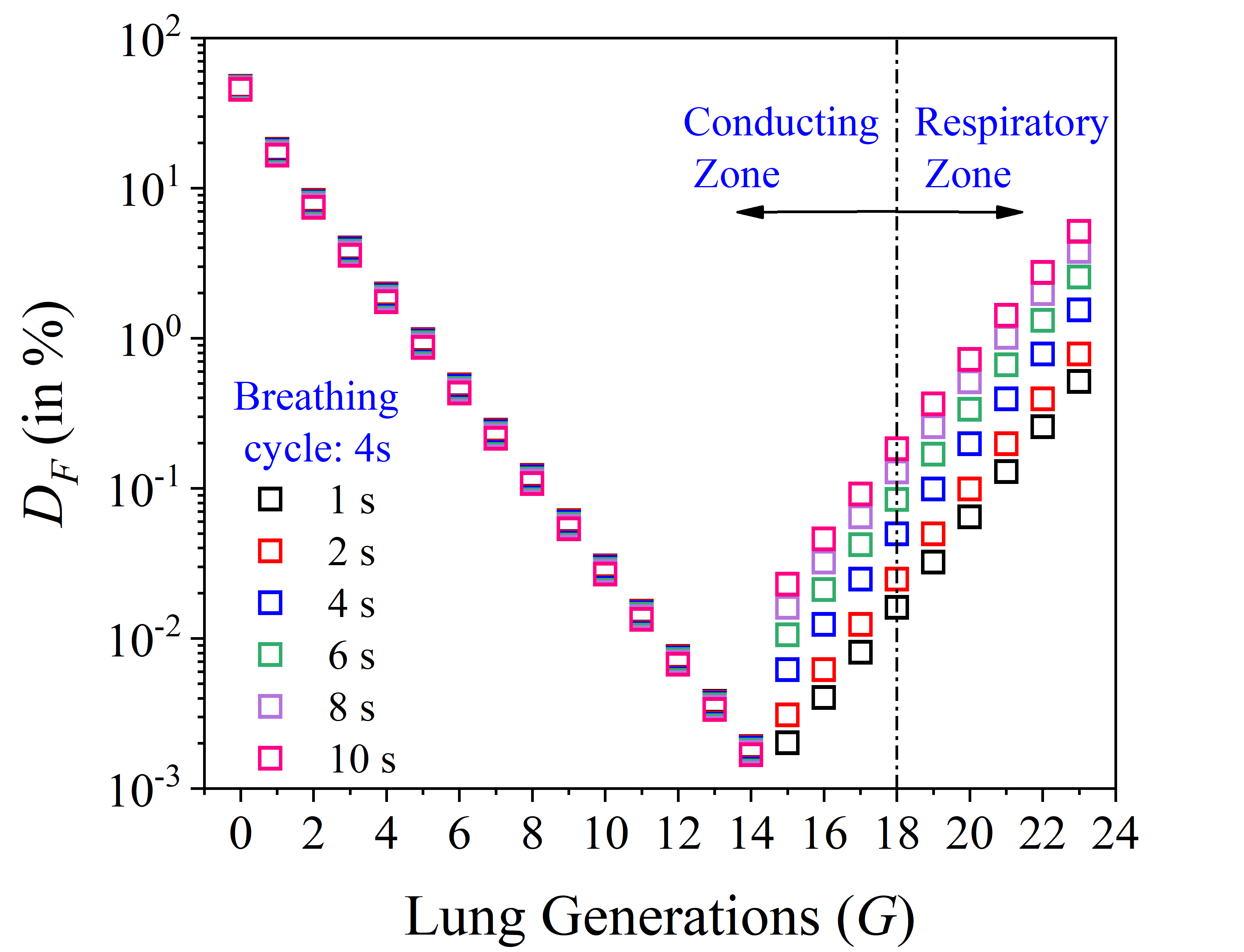}}
\caption{(a) Plot of $D_F$ and lung generations ($G$) for different breathing frequency. The time for each breathing cycle is considered as time for inspiration and expiration together. (b) Plot of $D_F$ and $G$ for different breath hold time. For both the plots it can be seen that model for $Re >> 1$ do not show any response with breathing frequency and breath hold time, whereas the model for $Re << 1$ found to be very sensitive for both of these parameters.}
\label{Figure: Effect of breathing frequency}
\end{figure}
\begin{figure}[hbt!]
\centering
{\includegraphics[width=12cm,height=11cm]{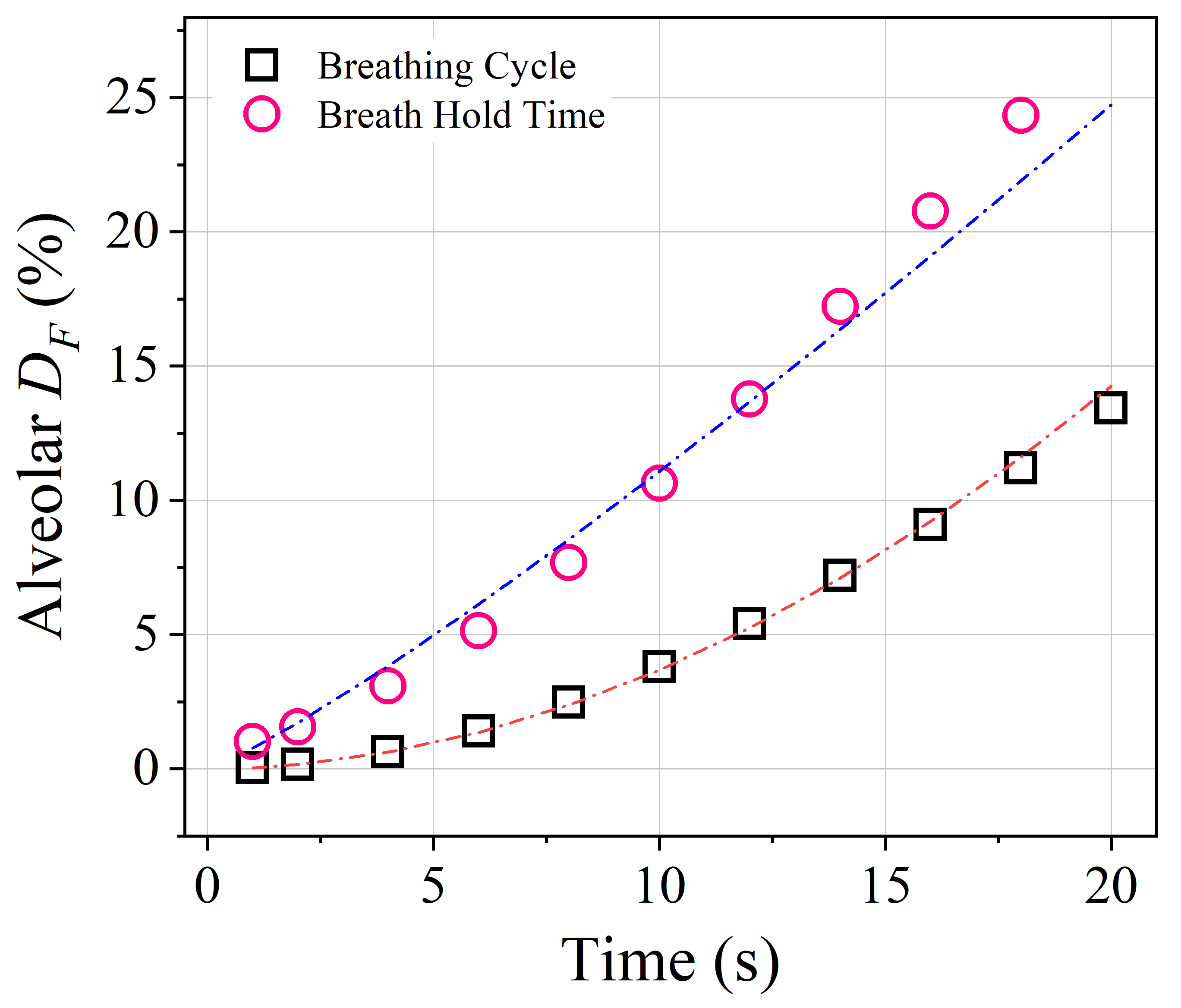}}

\caption{Plot of variation of alveolar deposition fraction ($D_F$) with time for different breathing frequency and breath hold time. The alveolar $D_F$ scales to $T^2$ for breathing cycle (Blue dotted line: $D_F = 0.04 T^{1.96}$ and scale to $T$ for breath hold time (Red dotted line: $0.766 T^{1.16}$)}
\label{Figure: Alveolar DF}
\end{figure}

The treatment of respiratory disease using nebulized drug finds its effectiveness when the drug deposits in the alveolar region of the lungs where gaseous exchanges takes place. The main goal of this study lies in estimating the regional deposition in the entire lung and increasing the amount of deposition in the alveolar region to increase drug efficacy. In this context, it is well-known that the breathing frequency and the breath hold time has a significant effect on aerosol deposition. The developed model is used to analyse the effect of breathing frequency and breath hold time for deposition of $10 \mu m$ sized particles, for single breathing cycle. Figure \ref{Figure: Effect of breathing frequency}(a) shows the effect of breathing frequency on regional deposition in lungs. The regional deposition fraction is analysed for different duration of breathing cycle (inspiration and expiration) ranging from $1s - 14s$. The variation of breathing frequency is mainly due to different types of activities we do round the day. The breathing cycle of 1s (inspiration: 0.5s and expiration:0.5s) takes place at the time of intense activity like running, swimming, cycling, climbing and other different workouts. The breathing cycle of 2s represents moderate activities like walking, light work outs etc and the cycle of 4s represents sedentary activities like sitting, lying down etc, which also represents the normal breathing cycle of human adult. From the figure it can be seen that for intense activity the deposition in the alveolar region region is significantly low and the total deposition in lungs is decreased. With a decrease in breathing frequency the deposition for $Re < 1$ increases whereas the deposition for $Re > 1$ is constant. This is because the model for $Re << 1$ (ref: equation \ref{eq: Best Fit:main}) indicates that the dimensionless deposition is proportional to $Re^{-2}$ whereas for $Re >> 1$, the deposition is independent of $Re$. Thus lower breathing frequency increases deposition in the alveolar region of the lungs due to high residence time of the aerosol which enhances the diffusion process and is desirable for high drug efficiency. The alveolar deposition can be further increased by introducing the breath hold time between inspiration and expiration. The figure \ref{Figure: Effect of breathing frequency}(b) shows the effect of breath hold time on alveolar deposition for breathing cycle of 4s and particle diameter of $10 \mu m$. The breath hold time contributes to the diffusion deposition process ($Re < 1$) which is dominant in the alveolar region in turn increasing aerosol deposition in the alveolus. The figure \ref{Figure: Alveolar DF} shows that the increase in alveolar deposition is non-linear with breathing frequency ($D_F \sim T^{2}$). Thus the increase of alveolar deposition becomes insignificant after certain duration of breathing cycle. But the alveolar deposition linearly increases with the breath hold time ($D_F \sim T^{1.2}$) which indicates that longer breath hold time can increase the chances of virus infection significantly and at the same time it is desirable for higher drug deposition. The figure \ref{Figure: Alveolar DF} shows that the increase in alveolar deposition is non-linear with breathing frequency ($D_F \sim T^{1.96}$). Thus the increase of alveolar deposition becomes insignificant after certain duration of breathing cycle. But the alveolar deposition linearly increases with the breath hold time ($D_F \sim T^{1.16}$) which indicates that longer breath hold time can increase the effects of drugs significantly. The effect of breath hold time is more significant than that of breathing frequency. Since both the curves are diverging with time (ref: figure \ref{Figure: Alveolar DF}), it can be inferred that higher breath hold time can be dangerous in terms of getting infected from the virus laden droplets. The addition of lower breathing frequency (i.e increase in inspiration and expiration time) can also cause the situation to worsen as it increases the residence time of the infected particles in the distal lung airways. The increase in the residence time will enhance the deposition in the alveolar airways since the diffusive deposition, which is the dominant mechanism in the distal lung, is directly proportional to time. Therefore, in addition to a lower breathing frequency, introduction of breath hold time in between inhalation and exhalation can increase the threat of virus infection in a crowded place. On other hand, in addition to a lower breathing frequency, breath hold time can be introduced in between inhalation and exhalation to maximise alveolar deposition and thus increase drug efficacy.

\clearpage

\par
In conclusion, we have developed a quantitative physics-consistent correlation to predict the rate of deposition in any bronchiole of the lung. This model can now become the building block to developing a model for the total and regional deposition in the lung. One could construct a branching tree structure of the distal branching structure and model total deposition as a sum of all the  bronchiole deposition. This modeling approach is grounded in experiments and could be construed to be complementary to purely computational whole lung simulation \citep{koullapis2020towards} approaches that are being pursued in the recent literature.


\chapter{Conclusion and Future Scope}
This chapter briefly lists out the key observations, understandings and conclusions from this study. It also discusses the future scope of this work which will provide deeper understanding of this study in order to improve the deposition model.
\section{Conclusion}
An experimental study of aerosol deposition in phantom bronchioles has been carried out for a wide range of Reynolds numbers ($10^{-2} < Re< 10^3$) and different bronchiole diameters ranging from $0.3 mm - 2 mm$ as well as for differing aspect ratios of the bronchiole. The experiments closely replicate flow and aerosol characteristics in the different generations of bronchioles in the lung. The aerosols were generated using an ultrasonic nebulizer, producing a mean droplet size $6.5 \mu m$. The aerosol particles were doped with boron quantum dots, the deposition of which was quantified by spectrofluorometer. The following conclusions can be drawn from the study:

\begin{enumerate}
    \item The dimensionless deposition in a particular bronchiole ($\delta$) is inversely proportional to the aspect ratio of the bronchiole ($\overline{L}$) (ref. figure \ref{Figure: Constant D}) but the effect of $Re$ diminishes with increasing $\overline{L}$.
    
    \item The value of $\delta$ increases exponentially with an increase in the dimensionless diameter ($\overline{D}$) for different $\overline{L}$. In addition, $\delta$ decreases with an increase in $\overline{L}$ for all $\overline{D}$. However, the variation of $\delta$ with $\overline{L}$ is small compared to its variation with $\overline{D}$.
    \item The value of $\delta$ is independent of the particle size based Reynolds number ($Re \overline{D}$). For all $Re \overline{D}$, $\delta$ exhibits an exponential increase with $\overline{D}$.
    \item $\delta \overline{L}$ is independent of $\overline{L}$ over several orders of magnitude of $Re$, which confirms the inverse relation between $\delta$ and $\overline{L}$ in figure \ref{Figure: Constant D}. 
    \item For low $Re$, $\delta\overline{L} \sim Re^{-2}$, indicating that the amount of aerosol deposited is independent of the flow conditions and only depends of the aerosol conditions. This is the case with \textit{diffusion} dominated deposition.
    \item The parameter regime where $10^{-1} < Re < 10$ is identified as the zone where \textit{sedimentation} is dominant.
    \item $\delta \overline{L}$ is independent of $Re$ for $10 < Re < 10^3$ which is identified as the \textit{impaction} regime. 
    \item The deposition model is developed using the experimental data (see equation \ref{eq: Best Fit:main}). The model predicts deposition fraction of 81.88 \% in the conducting airways ($0 \leqslant G \leqslant 17$) and 0.624 \% in the respiratory airways ($18 \leqslant G \leqslant 23$) for particle size $\sim \mathcal{O}(10 \mu m)$ which matches well with the results of \cite{hinds1999aerosol}.
    \item The breathing frequency and the breath hold time have significant effect on alveolar deposition. The drug delivery efficiency can be increased with low breathing frequency and by introducing high breath hold time between inspiration and expiration.
\end{enumerate}

\section{Future scope}
 This work is a fundamental deposition study that leads to understand the combined effects of different parameters influencing aerosol deposition in straight tubes. The quantitative analysis of deposition in micro capillaries is a challenging task which is addressed in this work by using Boron quantum dots as fluorescence aerosol. Although this study reports deposition in alveolar dimension capillaries with wide ranges of $Re$ mimicking the whole lungs, still there exists lots of future scopes for this work. This study can be extended in future in the following ways:
 \begin{enumerate}
    \item This study is performed with straight tubes for wide ranges of diameters. But the bifurcations of the lung bronchioles may play a significant role in deposition. Thus experiments can be carried out with "Y - shape" bifurcated tubes in future for development of better model.
    \item This study is restricted to suction of aerosol at a particular rate. The experimental setup can be modified to suck in and pump out mode as occurs in real breathing condition.
    \item The effect of gravity is not studied in this work although they affects the deposition process \citep{GoldbergSmith}. Thus the effect of different parameters can be studied in both horizontal and vertical capillaries to make the deposition model more rigid.
    \item This study is done with poly-dispersed aerosol with mean size of $6.5 \mu m$ which is common with commercially available nebulizers. Thus further investigations can be carried out with mono-dispersed particles of different sizes to study the effect of particle size on deposition, explicitly.
\end{enumerate}
 

\appendix

\chapter{MATLAB code for global pdf}
\label{appA}
clc\\
clear all \\

\% read data

str1 = 'Nebulizer'\\
Nrad =7;\\
width=2;\\
nbin=100;\\

for k=1:Nrad\\
    newstr = sprintf('\%.1d',k-1);\\
    Filename   =  strcat(str1, newstr,'.csv');\\
    dia{k,1}   =  xlsread(Filename,'C:C');\\
    vel{k,1}   =  xlsread(Filename,'B:B');\\
end\\

\%\% area calculation\\
rstart = -((Nrad-1)/2)*width;\\
rend   =  ((Nrad-1)/2)*width;\\

for i=1:Nrad\\
    radloc(i)=rstart+(i-1)*width;\\
end\\

for i=1:Nrad\\
    area(i)=1/2*abs(pi*(radloc(i)+width/2)$\wedge$2-pi*(radloc(i)-width/2)$\wedge$2);\\
end\\
area(((Nrad-1)/2)+1)=pi*(width/2)$\wedge$2;    \% Area at center location\\

\%\% max \& min size and velocity calculation to set axis\\

maxdrop=0;\\
mindrop=1000;\\
globalD=zeros(1,1);\\
globalA=zeros(1,1);\\
for i=1:Nrad\\
    temp = dia{i,1};\\
    maxdrop1=max(temp);\\
    maxdrop=max(maxdrop1,maxdrop);\\
    mindrop1=min(temp);\\
    mindrop=min(mindrop1,mindrop); \\
end\\
$[$ sg dummy $]$ = size(globalD);\\
globalD=globalD(2:sg,1);\\
globalA=globalA(2:sg,1);\\
globalA=int16(globalA);\\

clear temp\\

maxvel=0;\\
minvel=100;\\
globalV=zeros(1,1);\\
for i=1:Nrad\\
    temp    = vel{i,1};\\
    maxvel1=max(temp);\\
    maxvel=max(maxvel1,maxvel);\\
    minvel1=min(temp);\\
    minvel=min(minvel1,minvel); \\      
end\\
$[$ sg dummy $]$ =size(globalV);\\
globalV=globalV(2:sg,1);\\

clear temp\\

\%\% bining method for size and velocity pfd calculation\\

grat=(maxdrop/mindrop)$\wedge$(1/nbin);\\
X=mindrop*grat.$\wedge$(0:nbin);\\
X(1)=mindrop*0.5; \% Geometric progression of bins\\
X(nbin+1)=maxdrop*1.1;\\
 
\% A=(1.02*maxdrop-0.5*mindrop)/nbin;\\
 B=(1.02*maxvel-1.01*minvel)/nbin;\\
\% X=[0.5*mindrop:A:maxdrop*1.02]; \%Arithematic progression of bins\\
Y=[1.01*minvel:B:maxvel*1.02];\\

\%\% diameter pdf calculation \\

N1=zeros(1,nbin+1);\\
N2=zeros(1,nbin+1);\\
for i=1:Nrad\\
    temp    = dia{i,1};\\
   $[$ stemp dummy $]$ =size(temp);\\
    temp1=temp(:,1);   \\                
    $[$ N1,y1 $]$ =hist(temp1,X);\\
    N2=N1*area(i)+N2;\\
end\\
clear temp\\
clear temp1\\

N2=N2/sum(N2);\\
normfac=trapz(X,N2);\\
N2=N2/normfac;\\

\%\% velocity pdf calculation \\

N3=zeros(1,nbin+1);\\
N4=zeros(1,nbin+1);\\
for i=1:Nrad\\
    temp    = vel{i,1};\\
   $[$ stemp dummy $]$ =size(temp);\\
    temp1=temp(:,1);   \\                 
    $[$ N3,y1 $]$ =hist(temp1,Y);\\
    N4=N3*area(i)+N4;\\
end\\

N4=N4/sum(N4);\\
normfac=trapz(Y,N4);\\
N4=N4/normfac;\\

\%\%\\
figure(1)\\
scrsz = get(0,'ScreenSize');\\
h = gcf;\\
set(h,'Position',[0 0 2000 1000]);\\
plot(X,N2,'ks','LineWidth',2.5,'MarkerEdgeColor','k','MarkerSize',15)\\
xlabel(' $x$','Interpreter','latex','FontWeight','bold','FontSize',35,'FontName','Times new romen');\\
ylabel('$\mathcal{X}(x)$','interpreter','latex','rot',90,'FontWeight','bold','FontSize',35,'FontName','Times new romen');\\

\%\%\\
figure(2)\\
scrsz = get(0,'ScreenSize');\\
h = gcf;\\
set(h,'Position',[0 0 2000 1000]);\\
plot(Y,N4,'ks','LineWidth',2.5,'MarkerEdgeColor','k','MarkerSize',15)\\
xlabel(' $u$','Interpreter','latex','FontWeight','bold','FontSize',35,'FontName','Times new romen');\\
ylabel('$\mathcal{U}(u)$','interpreter','latex','rot',90,'FontWeight','bold','FontSize',35,'FontName','Times new romen');

hold off;\\

\begin{singlespace}
  \bibliography{refs}
\end{singlespace}


\listofpapers

\begin{enumerate}  
\item Mallik AK, Mukherjee S, Panchagnula MV. An experimental study of respiratory aerosol transport in phantom lung bronchioles. Physics of Fluids. 2020 Nov 1;32(11):111903.
\item Mallik, A.K., Sarma, T.P., Roy, A., Panchagnula, M.V. and Seshadri, S., 2020. PHASE DOPPLER PARTICLE ANALYSER (PDPA) CHARACTERIZATION AND MODELING OF SPRAYS FROM ORTHOGONALLY INTERACTING WATER AND AIR JETS. Journal of Flow Visualization and Image Processing, 27(2).
\end{enumerate}  

\end{document}